\documentclass{article}

\usepackage{float}
\usepackage{PRIMEarxiv}
\usepackage{siunitx}
\usepackage{amsmath} 
\usepackage[utf8]{inputenc} 
\usepackage[T1]{fontenc}    
\usepackage{hyperref}       
\usepackage{url}            
\usepackage{booktabs}       
\usepackage{amsfonts}       
\usepackage{nicefrac}       
\usepackage{lettrine}       
\usepackage{microtype}      
\usepackage{lipsum}
\usepackage{fancyhdr}       
\usepackage{graphicx}       
\graphicspath{{media/}}     

\title{Time-Lagged Recurrence: a data-driven method to estimate the predictability of dynamical systems}

\author{
  Chenyu Dong \\
  College of Design and Engineering \\
  National University of Singapore \\
  \texttt{chenyu.dong@u.nus.edu} \\
  \And
  Davide Faranda \\
  LSCE, Université Paris-Saclay, France \\
  London Mathematical Laboratory, UK \\
  IPSL, École Normale Supérieure, France \\
  \texttt{davide.faranda@lsce.ipsl.fr} \\
  \AND
  Adriano Gualandi \\
  Department of Earth Sciences \\
  University of Cambridge \\
  Istituto Nazionale di Geofisica e Vulcanologia \\
  \texttt{ag2347@cam.ac.uk} \\
  \AND
  Valerio Lucarini \\
  School of Computing and Mathematical Sciences \\
  University of Leicester, UK \\
  \texttt{v.lucarini@leicester.ac.uk} \\
  \And
  Gianmarco Mengaldo\textsuperscript{*} \\
  College of Design and Engineering \\
  National University of Singapore \\
  \texttt{mpegim@nus.edu.sg} \\
}

\begin{document}
\maketitle
\begin{abstract}
Nonlinear dynamical systems are ubiquitous in nature and they are hard to forecast. Not only they may be sensitive to small perturbations in their initial conditions, but they are often composed of processes acting at multiple scales. 
Classical approaches based on the Lyapunov spectrum rely on the knowledge of the dynamic forward operator, or of a data-derived approximation of it. 
This operator is typically unknown, or the data are too noisy to derive its faithful representation. 
Here we propose a new data-driven approach to analyze the local predictability of dynamical systems. 
This method, based on the concept of recurrence, is closely linked to the well-established framework of local dynamical indices. 
When applied to both idealized systems and real-world datasets arising from large-scale atmospheric fields, our new approach proves its effectiveness in estimating local predictability. 
Additionally, we discuss its relationship with other local dynamical indices, and how it reveals the scale-dependent nature of predictability.
Furthermore, we explore its link to information theory, its extension that includes a weighting strategy, and its real-time application. 
We believe these aspects collectively demonstrate its potential as a powerful diagnostic tool for complex systems.
\end{abstract}


\section{Introduction}\label{main-introduction}
Systems that evolve in time according to specific rules are known as \textit{dynamical systems}, and the associated field of study is called \textit{dynamical system theory}. 
There is a rich history on this topic, starting from the 1600s, when Newton developed calculus and classical mechanics~\cite{newton1833philosophiae}, to the 1800s, when Poincar\'e developed geometrical approaches to the study of dynamical systems~\cite{poincare1890probleme}. 
In the 1900s, the field flourished, with the pioneering contributions of Birkhoff, Kolmogorov, Moser, Arnold, Lorenz, Ruelle, Takens, May, and Feigenbaum, among others, who laid the foundations of modern dynamical system theory, including works on oscillators, chaos and fractals, well summarized by Strogatz~\cite{strogatz2018nonlinear}.

Many real-world systems are made of multiple interacting parts, giving rise to complex dynamical systems. 
A paradigm for describing such systems relies on considering high-dimensional deterministic models, which are often characterised by chaotic behaviour and multiscale dynamics~\cite{eckmann1985ergodic,pikovsky2016lyapunov}. 
A separate paradigm relies on considering stochastic dynamics, where randomness is explicitly introduced in the system \cite{Risken2012,Pavliotis2014}. 
When constructing reduced order models of high dimensional systems, stochastic dynamics is an emergent property. 
This is at the core of the Hasselmann's program for climate science~\cite{Hasselmann1976,LC2023} and can be rigorously justified thanks to the Mori-Zwanzig theory~\cite{Mori1965,Zwanzig1961}.
The latter also lays the foundation for the use of data-driven methods aimed at estimating the role of the unresolved scales of motion on the resolved ones \cite{Santos2021}. 
And, indeed, low-dimensional stochastic models have proved very valuable tools for understanding the statistical and dynamical properties of complex systems, e.g. in the area of Earth system science \cite{Penland1993,Kondrashov2006,faranda2017stochastic,gualandi2023deterministic,nath2009ground,Giorgini2022}.

Either way, a well-known fact is that we currently observe a finite predictability horizon for systems of interest like, for example, the atmosphere~\cite{lorenz1963deterministic,siegert2016prediction}.
This means that, when forecasting the future behaviour of a dynamical system, the error that we make tends to increase the longer into the future we want to forecast~\cite{palmer1993extended}. 

To understand predictability and the methods commonly used to assess it, we look at the evolution of the system in the so-called phase space, that is the space of all the variables used to describe it.
In the phase space, usually we can identify a subset, namely a finite-size manifold of trajectories followed by the dynamical system over time and denoted by ${\bf{x}}(t)$, where ${\bf{x}}$ is the vector of observables that depends on time $t$. 
This finite-size manifold is also known as attractor (random attractor in the case of stochastic dynamical systems~\cite{crauel1997random}).

The origin of this practical unpredictability may stem either from an intrinsic stochastic nature of the system or from the ignorance of the current state of the system and/or of the dynamic rule controlling its evolution~\cite{boffetta2002predictability}, or from a combination of both. 
Furthermore, it is well-known that even low-dimensional nonlinear systems may be difficult to forecast if they exhibit an exponential divergence of close-by states~\cite{lorenz1963deterministic}.
In such cases the unpredictability comes from our lack of precise knowledge of the current state of the system: if we knew exactly the initial conditions we would always get the same future evolution.
A common quantity used to characterize the predictability of a system is the maximum Lyapunov exponent $\lambda_{\rm{max}}$, which measures the fastest possible rate of exponential divergence between two nearby trajectories in the phase space for an infinitely long time horizon.
If we look at the evolution of two trajectories starting from two points close to each other, namely ${\bf{x}}(t_0)$ and ${\bf{x}}(t_0) + {\delta}{\bf{x}}$, these will diverge exponentially fast as time progresses, until their distance will plateau around a value bounded by the diameter of the attractor~\cite{parlitz2016estimating}. 
Even more instructive is to look at the evolution of a ball around the state ${\bf{x}}(t)$.
From this, one can define a full Lyapunov spectrum, made of Lyapunov exponents equal in number to the variables used to describe the phase space.
The Lyapunov exponents are mean logarithmic growth rates of the lengths of the principal axes of this ball evolving under the action of the dynamic rule~\cite{oseledec1968multiplicative,eckmann1985ergodic,parlitz2016estimating,pikovsky2016lyapunov}.
The existence of at least one positive Lyapunov exponent is the hallmark of chaotic dynamical systems. 
Therefore, measuring predictability in dynamical systems can be re-framed as computing the divergence of a system's nearby trajectories over time. 

It is important to highlight that the Lyapunov exponents are averaged quantities and the average is typically performed over a long trajectory.
As a consequence, they serve as key indicators of the system's global predictability~\cite{oseledec1968multiplicative,eckmann1985ergodic}, where the term \textit{global} refers to average properties of the system, hence to the average predictability of the system. 
While the average predictability of the system can be useful, theoretical curiosity as well as operational requirements and priorities indicate the need to investigate a local notion of predictability. 
Indeed, the stability properties of a dynamical system can be dramatically state-dependent \cite{Maiocchi2023}. 
We then wish to understand the system's predictability in a specific region of the phase space (we will refer to these local characteristics also as \textit{instantaneous}, since they refer to a specific time)~\cite{nese1989quantifying,ziehmann2000localized}. This notion has extreme relevance in the context of weather prediction \cite{Slingo2011} and key implications for devising efficient and accurate data assimilation methods \cite{Carrassi2018}.

To this end, local (or finite-time) Lyapunov exponents~\cite{abarbanel1991variation,eckhardt1993local,ziehmann2000localized,Cencini2013} are useful quantities.
However, their reliance on linearized dynamics prevents the characterization of the system outside the linear regime of perturbations. 
Acknowledging this limitation, finite-size Lyapunov exponents~\cite{aurell1997predictability,boffetta1998extension} and nonlinear local Lyapunov exponents~\cite{ding2007nonlinear,huai2017quantifying} were subsequently proposed to quantify local predictability in the nonlinear regime, as they consider finite perturbations. 
Nonetheless, it is noteworthy that they both assume an exponential law for error growth. 
Although this is consistent with classic Lyapunov exponents within the linear regime, this assumption often does not hold in the nonlinear regime (i.e., for finite errors).
These local indices have been applied successfully to various dynamical systems of varied complexity to gain insights into their predictability~\cite{ding2007nonlinear,geist1990comparison,aurell1997predictability,eckhardt1993local}, despite the aforementioned limitations.

Besides Lyapunov exponents and their variants, information theory also offers methods to quantify predictability~\cite{delsole2007predictability,boffetta2002predictability}. 
Within this framework, various entropy-based indicators are usually employed as predictability measures for model-based prediction, with varying degrees of success~\cite{schneider1999conceptual, kleeman2007statistical,leung1990information, kleeman2002measuring}.
However, because these methods are model-based, they are unsuitable for observational datasets.
Other indices from nonlinear time series analysis also use predictability measures such as sample entropy, permutation entropy, and others~\cite{li2007predictability,ikuyajolu2021information,bandt2002permutation}; nevertheless, they are only effective with time series or systems with relatively low dimensionality.
To the best of the authors' knowledge, no information theory method can infer local predictability information from high-dimensional observational data.

From the brief review of relevant literature just outlined, the above mentioned indices have several drawbacks. 
More specifically, Lyapunov exponents and their variants may not work well for nonlinear regimes, while information-theory-based methods may not be directly applicable to high-dimensional dynamical systems. 
In the next section, we leverage the recent framework proposed by Lucarini et al.~\cite{lucarini2016extremes}, which uses recurrences to characterize local dynamical properties, to develop a novel local predictability index that differs from existing metrics, such as the Lyapunov exponents and their variants.

\section*{A local data-driven approach}
\label{sec:index}

Lucarini et al.,~\cite{lucarini2016extremes} introduced a novel framework to measure the local properties of dynamical systems. In particular, they combine extreme value theory~\cite{fisher1928limiting} and the Poincaré recurrence theorem~\cite{poincare1890probleme} to construct purely data-driven local dynamical indices, that can measure the local dimension of dynamical systems and their persistence at any point in time. 
Under the ergodic assumption, the Poincaré recurrence theorem guarantees that the system will visit an arbitrarily small neighborhood of any state again and again as time passes.
In this framework, for one given state of interest ${\boldsymbol{\zeta}} = {\bf{x}}(t={t}_{\boldsymbol{\zeta}})$ in the phase space, its neighboring states, also known as \textit{recurrences}, are used to characterize its local dynamical properties. 
This can be formalized by defining the negative logarithmic distance between ${\boldsymbol{\zeta}}$ and all other states
\begin{equation}\label{eq:log_dist}
g\left({\boldsymbol{\zeta}}, {\bf{x}}(t)\right) = -\log \left[\operatorname{dist}\left({\bf{x}}(t), {\boldsymbol{\zeta}}\right)\right],
\end{equation}
where the dist function can be any distance metric, and by 
subsequently defining a quantity called \textit{exceedance} for the neighboring states (i.e., recurrences): 
\begin{equation}\label{eq:exceedance}
{\bf{u}}({\boldsymbol{\zeta}}) = g({\boldsymbol{\zeta}}, {\bf{x}}(t)) - s(q, {\boldsymbol{\zeta}}), \;\;\;\; \forall ~g({\boldsymbol{\zeta}}, {\bf{x}}(t)) > s(q, {\boldsymbol{\zeta}}).
\end{equation}
In Eq.~\ref{eq:exceedance}, $s(q, {\boldsymbol{\zeta}})$ is a high threshold corresponding to the $q$-quantile of the series $g\left({\boldsymbol{\zeta}}, {\bf{x}}(t)\right)$, noting that having $g({\boldsymbol{\zeta}}, {\bf{x}}(t))$ above the threshold $s(q, {\boldsymbol{\zeta}})$ is a synonym of requiring a trajectory to fall within a neighborhood of ${\boldsymbol{\zeta}}$ (intended as a ball).

Building on the above-defined quantities and leveraging extreme value theory, this framework introduces two local dynamical indices: the local dimension $d$ and persistence $\Theta$ (see Materials and Methods for a detailed explanation and derivation), both of which are frequently associated with the predictability of dynamical systems~\cite{gualandi2020predictable,faranda2017dynamical,hochman2019new}.
More specifically, high local dimension and low persistence are seen as an indication that the system is less predictable. 
The rationale for this is as follows. 
If a system requires more degrees of freedom (high $d$) to be described at a given point in the phase space, then its complexity is greater at that given point, making it possibly less predictable.
Similarly, if a system leaves a neighborhood quickly (low $\Theta$), then the dynamics in that neighborhood are fast and non-persistent, an aspect that may indicate reduced predictability.
While these arguments on predictability are qualitatively valid, and frequently correct, they approach predictability from a complexity ($d$), and mean residence time ($\Theta$) perspective, without directly addressing predictability.

\begin{figure}[h!]
\centering
\includegraphics[width=0.85\columnwidth]{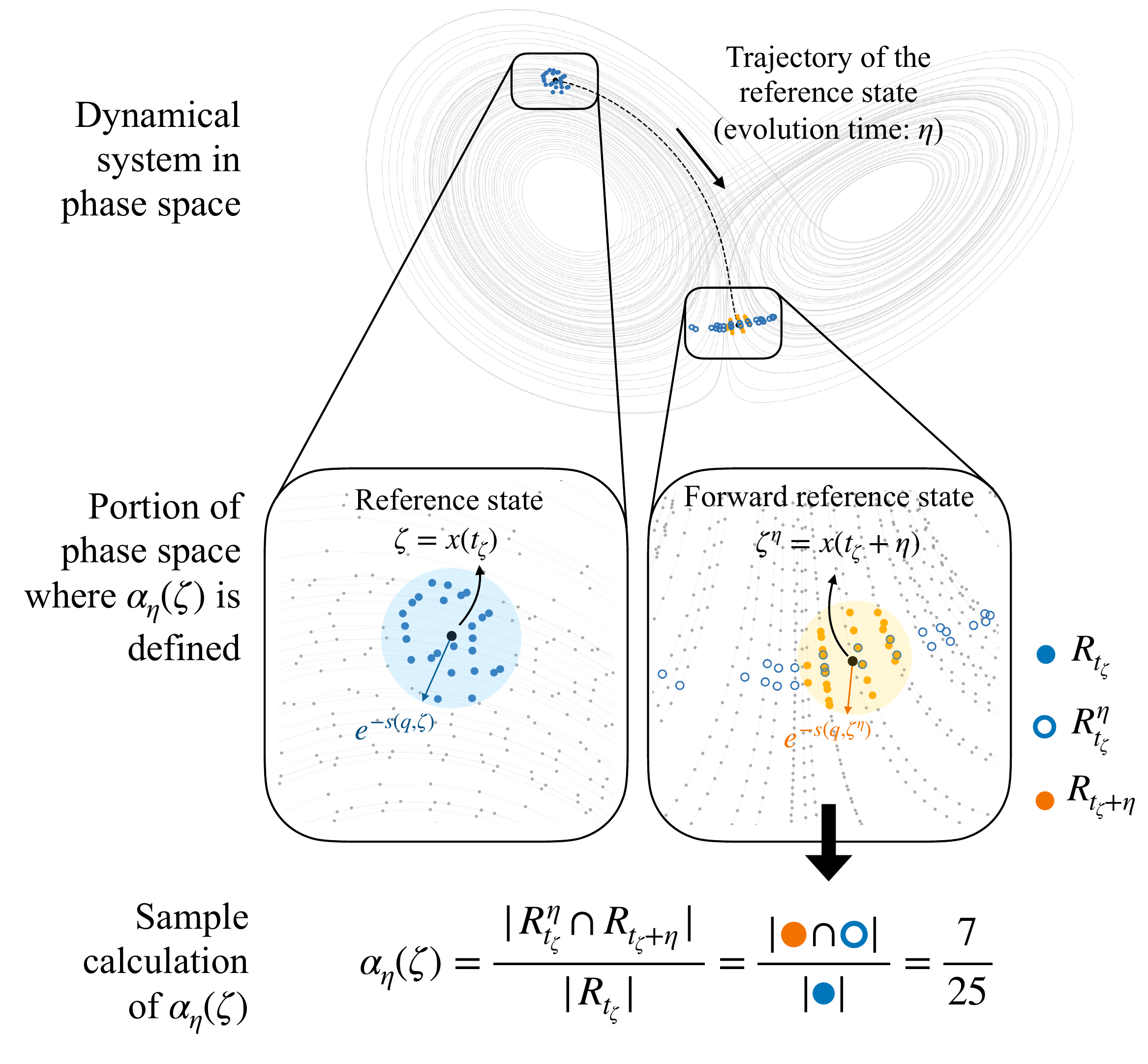}
\caption{Schematic illustration of the computation of $\alpha_{\eta}(t)$, demonstrated in the phase space of the Lorenz-63 system. This figure presents all steps involved in computing the \textit{time-lagged recurrence} for the reference state $\zeta$ at a forecasting horizon $\eta$, namely $\alpha_{\eta}(\zeta)$. The panels in the second row provide a zoomed-in view of the phase space region where $\alpha_{\eta}(\zeta)$ is defined. \textit{Recurrences} ($R_{t_{\zeta}}$), \textit{forward recurrences} ($R_{t_{\zeta}}^{\eta}$), and \textit{forward-reference-state recurrences} ($R_{t_{\zeta}+{\eta}}$) are represented by solid blue dots, empty blue dots, and orange dots, respectively. The blue circle with radius $e^{-s(q, \zeta)}$ indicates the hypersphere used to define the neighborhood of the reference state, while the orange circle with radius $e^{-s(q, \zeta^\eta)}$ corresponds to the forward-reference-state.}
\label{fig:fig1}
\end{figure}

In this work, we address the predictability of dynamical systems by building upon the framework of local dynamical indices just outlined. 
In particular, we introduce a new quantity named `\textit{time-lagged recurrence}' and denoted by $\alpha_{\eta}$. 
This new quantity uses neighboring states to monitor an ensemble of trajectories, and provides an indication in a statistical sense of the local predictability of dynamical systems. This is accomplished in a purely data-driven fashion, i.e. without requiring a model nor further simplifications on the data. 

The computation of $\alpha_{\eta}$ also relies on the concept of recurrences and the quantities defined above in Eqs.~\ref{eq:log_dist} and~\ref{eq:exceedance}, with a schematic illustration presented in Fig.~\ref{fig:fig1}.
More specifically, we identify the set of states -- i.e., recurrences -- $R_{t_{\boldsymbol{\zeta}}}$ in the neighborhood of the reference state $\boldsymbol{\zeta}$, that is:
\begin{equation}\label{eq:recurrences}
R_{t_{\boldsymbol{\zeta}}} = \left\{ {\bf{x}}(t_{u_i}) : u_i(\boldsymbol{\zeta}) > 0 \wedge |t_{\boldsymbol{\zeta}} - t_{u_i}| > w \right\}_{i=1}^{N_n},
\end{equation}
where $u_i(\boldsymbol{\zeta})$ represents the exceedances, defined in Eq.~\ref{eq:exceedance}, and with positive values indicating proximity to the reference state in phase space, $t_{\boldsymbol{\zeta}}$ is the time that identifies the reference state $\boldsymbol{\zeta}$, $N_n$ is the number of recurrences (i.e., the number of temporally non-adjacent neighbors) at time $t_{\boldsymbol{\zeta}}$, $t_{u_i}$ is the time corresponding to the exceedance $u_i$, and $w$ is the \textit{Theiler window}~\cite{theiler1986spurious} that excludes temporal neighbors of $\boldsymbol{\zeta}$.
We note that if $w$ is chosen too small that cannot exclude temporal neighbors it will lead to spurious high values of $\alpha_{\eta}$.

After having identified the \textit{recurrences} $R_{t_{\boldsymbol{\zeta}}}$, to characterise the local predictability of $\boldsymbol{\zeta}$, we first need to estimate how neighboring states evolve after time $\eta$ has passed from $t_{\boldsymbol{\zeta}}$. 
To this end, we first evolve the states within $R_{t_{\boldsymbol{\zeta}}}$ forward by time $\eta$ (where $\eta$ is also referred to as forecasting horizon), which leads to a set of new states, that we name \textit{forward recurrences} 
\begin{equation}\label{eq:forward_recurrences}
R_{t_{\boldsymbol{\zeta}}}^{\eta} = \left\{{\bf{x}}(t+\eta) \;\;\; \forall {\bf{x}}(t) \in R_{t_{\boldsymbol{\zeta}}} \right\}.
\end{equation}
These new states, that were evolved forward in time by $\eta$ from time $t_{\boldsymbol{\zeta}}$, are represented by the circles with blue contours in Fig.~\ref{fig:fig1}, and are denoted by using the superscript $(\cdot)^{\eta}$.

Once the forward recurrences have been defined (Eq.~\ref{eq:forward_recurrences}), we need to define the \textit{forward reference state}, that is: the reference state $\boldsymbol{\zeta} = {\bf{x}}(t_{\boldsymbol{\zeta}})$ after time $\eta$, and denoted by $\boldsymbol{\zeta}^{\eta} = {\bf{x}}(t_{\boldsymbol{\zeta}}+\eta)$. This forward reference state will have its own excedeences ${\bf{u}}({\boldsymbol{\zeta}}^{\eta})$ made of $N_n$ elements $\lbrace u_i(\boldsymbol{\zeta}^\eta)\rbrace_{i=1}^{N_n}$, and we can compute the \textit{forward-reference-state recurrences} 
\begin{equation}\label{eq:new_reference_recurrences}
R_{t_{\boldsymbol{\zeta}}+\eta} = \left\{ {\bf{x}}(t_{u_i}) : u_i(\boldsymbol{\zeta}^{\eta}) > 0 \wedge |t_{\zeta} +\eta - t_{u_i}| > w \right\}_{i=1}^{N_n},
\end{equation}
in analogy to what was done for the reference state $\boldsymbol{\zeta}$ in Eq.~\ref{eq:recurrences}.

With the three sets just introduced in Eqs.~\ref{eq:recurrences}--\ref{eq:new_reference_recurrences}, namely the recurrences $R_{t_{\boldsymbol{\zeta}}}$ in Eq.~\ref{eq:recurrences}, the forward recurrences $R_{t_{\boldsymbol{\zeta}}}^{\eta}$ in Eq.~\ref{eq:forward_recurrences}, and the forward-reference-state recurrences $R_{t_{\boldsymbol{\zeta}}+\eta}$ in Eq.~\ref{eq:new_reference_recurrences}, we can define the \textit{local predictability index} as follows:
\begin{equation}\label{eq:alphat}
\alpha_{\eta}(\boldsymbol{\zeta}) = \frac{|R_{t_{\boldsymbol{\zeta}}}^\eta \cap R_{{t_{\boldsymbol{\zeta}}+\eta}}|}{|R_{t_{\boldsymbol{\zeta}}}|}
\end{equation}
where $|(\cdot)|$ represents the number of elements within the set $(\cdot)$, and $\cap$ is the intersection of two sets. 
Since both $R_{t_{\boldsymbol{\zeta}}}$ and $R_{{t_{\boldsymbol{\zeta}}+\eta}}$ are defined using the same quantile-based thresholds, the number of elements in one set is equal to the number of elements in the other set, that is: $|R_{{t_{\boldsymbol{\zeta}}}}| = |R_{t_{\boldsymbol{\zeta}}+\eta}| = N_n$. 
In addition, given that $R_{t_{\boldsymbol{\zeta}}}^{\eta}$ is defined from $R_{t_{\boldsymbol{\zeta}}}$, the number of elements it contains is also equal to $N_n$. 
It follows that 
$0 \leq \alpha_{\eta}({\bf{x}}(t)) \leq 1$. 
A value of $\alpha_{\eta}(\boldsymbol{\zeta})$ close to 1 indicates high probability for a neighbor state to remain close to the reference state.
We interpret this as a high predictability, in a statistical sense, for state $\boldsymbol{\zeta}$ at time $t_{\boldsymbol{\zeta}}$, and forecasting horizon $\eta$, suggesting that most of its neighboring states stay close to the future trajectory of $\boldsymbol{\zeta}$ as time evolves. 
Conversely, when $\alpha_{\eta}(\boldsymbol{\zeta})$, is close to zero, this implies that most of the initial neighbors have moved away from the trajectory of $\boldsymbol{\zeta}$ after time $\eta$, indicating low predictability in a statistical sense. 

The derivation of the new quantity $\alpha_{\eta}(\boldsymbol{\zeta})$ relies on the distance function $\operatorname{dist}$ and the quantile $q$, both of which are used to define recurrences. Although the Euclidean distance has been shown to be an appropriate choice for physical systems~\cite{lucarini2016extremes}, the distance function can be adjusted depending on the prior knowledge of the system and the application context.
The selection of the quantile $q$ can be adapted for different application scenarios, and it reflects the local statistical predictability for different scales. This is not surprising as we are defining $\alpha_\eta$ using extreme value theory where quantile choice is always an important step. 
In SI Appendix section 2, we show an example of this scale-dependent behavior using various $q$.
This dependence on $q$ can be seen as a favorable property as it allows for analyzing predictability at different scales.

In addition, the new quantity $\alpha_{\eta}(\boldsymbol{\zeta})$ is inherently connected to information theory, since estimating local predictability can be viewed as a process that quantifies the uncertainty or randomness within the system. In fact, if we take a conditional probability perspective, where $P(A|B)$ represents the conditional probability of $A$ given $B$, then we can reformulate the definition of $\alpha_{\eta}$ in Eq.~\ref{eq:alphat} as follows:
\begin{equation}\label{eq:conditional}
\alpha_{\eta}(\boldsymbol{\zeta})=P[{\bf{x}}(t+\eta) \in R_{{t_{\boldsymbol{\zeta}}+\eta}} | {\bf{x}}(t) \in R_{t_{\boldsymbol{\zeta}}}],
\end{equation}
where $A = {\bf{x}}(t+\eta) \in R_{{t_{\boldsymbol{\zeta}}+\eta}}$ corresponds to recurrences of the reference state $\boldsymbol{\zeta}$ that, after being propagated forward in time by $\eta$, belong to the forward-state neighborhood $\boldsymbol{\zeta}^{\eta}$ (i.e., they are forward-reference-state
recurrences), and $B = {\bf{x}}(t) \in R_{t_{\boldsymbol{\zeta}}}$ corresponds to the recurrences of the reference state $\boldsymbol{\zeta}$. We can further link $\alpha_\eta$ to information theory using the Shannon entropy, when defined -- see SI Appendix section 3.

The new predictability index has a number of advantages compared to existing predictability measures: (i) it is purely data-driven, (ii) it is local as it measures instantaneous properties, (iii) it does not require simplifications to be made on the data, and (iv) it is suitable for high dimensional datasets. 

However, given that it uses future states, it cannot be readily used for real-time predictability analyses, for instance in the context of operational weather forecasting. 
We shall discuss this aspect in a later section and propose a practical way to address this limitation.

It is helpful here to emphasize some key differences to other data-driven ways of measuring local predictability properties. The classical algorithms proposed in~\cite{Wolf1985} to measure the Lyapunov exponents from available time series allow to estimate the growth rate of the $n-$dimensional volumes (for $n$ smaller than or equal to the dimension of the phase space) defined by $n+1$ initially nearby points. The strategy there strongly relies on using delay coordinates in order to take advantage of Takens' theorem~\cite{Takens1981} and reconstruct the key properties of the attractor. By selecting nearby points belonging to a region of interest, local properties can be extracted. 
Following such geometrical construction, one obtains that \begin{equation}\label{Lyapunovfinite}
    P[\mathrm{dist}(\bf{x}(t+\eta),\zeta^\eta) <\epsilon |  \mathrm{dist}(\bf{x}(t),\zeta) <\epsilon]\approx e^{-\left(\small{\sum}_{\lambda_j>0}\lambda_j^{\zeta}\right) \eta}
\end{equation}
where the exponent is proportional to the sum of positive local, finite-time, Lyapunov exponents. Clearly $\epsilon$ has to be small. 
After comparing Eqs.~\ref{eq:conditional} and~\ref{Lyapunovfinite} it is tempting to deduce $\alpha_{\eta}\approx e^{-(\small{\sum}_{\lambda_j>0}\lambda_j^{\zeta}) \eta}$ (see later discussion on specific numerical examples). Yet, a key difference between the two definitions exist when one applies them to actual data, because in Eq.~\ref{eq:conditional} one considers neighborhoods of $\zeta$ and $\zeta^\eta$ having the same number of points (thanks to the use of quantiles in defining the threshold of the distance observables), so that such neighborhoods will in general have a different radius (which is instead identical and equal to $\epsilon$ in Eq.~\ref{Lyapunovfinite}).

\section*{Results and discussion}
\label{sec:results}

We have computed the new metric $\alpha_{\eta}(\boldsymbol{\zeta})$ for systems of varying complexity to show its effectiveness. The following subsections provide a detailed discussion of the results for the Lorenz-63 attractor and climate data characterizing atmospheric circulation in the Euro-Atlantic sector. Additionally, a brief summary of $\alpha_{\eta}(\boldsymbol{\zeta})$ applied to other systems is presented in a subsequent section to illustrate its broad applicability.
Following the discussion of results obtained from various systems, we propose a method for utilizing this index in real-time, addressing the issue introduced at the end of the previous section.

\subsection*{Lorenz-63}\label{subsec:idealized}
\begin{figure}[h!]
\centering
\includegraphics[width=0.8\columnwidth]{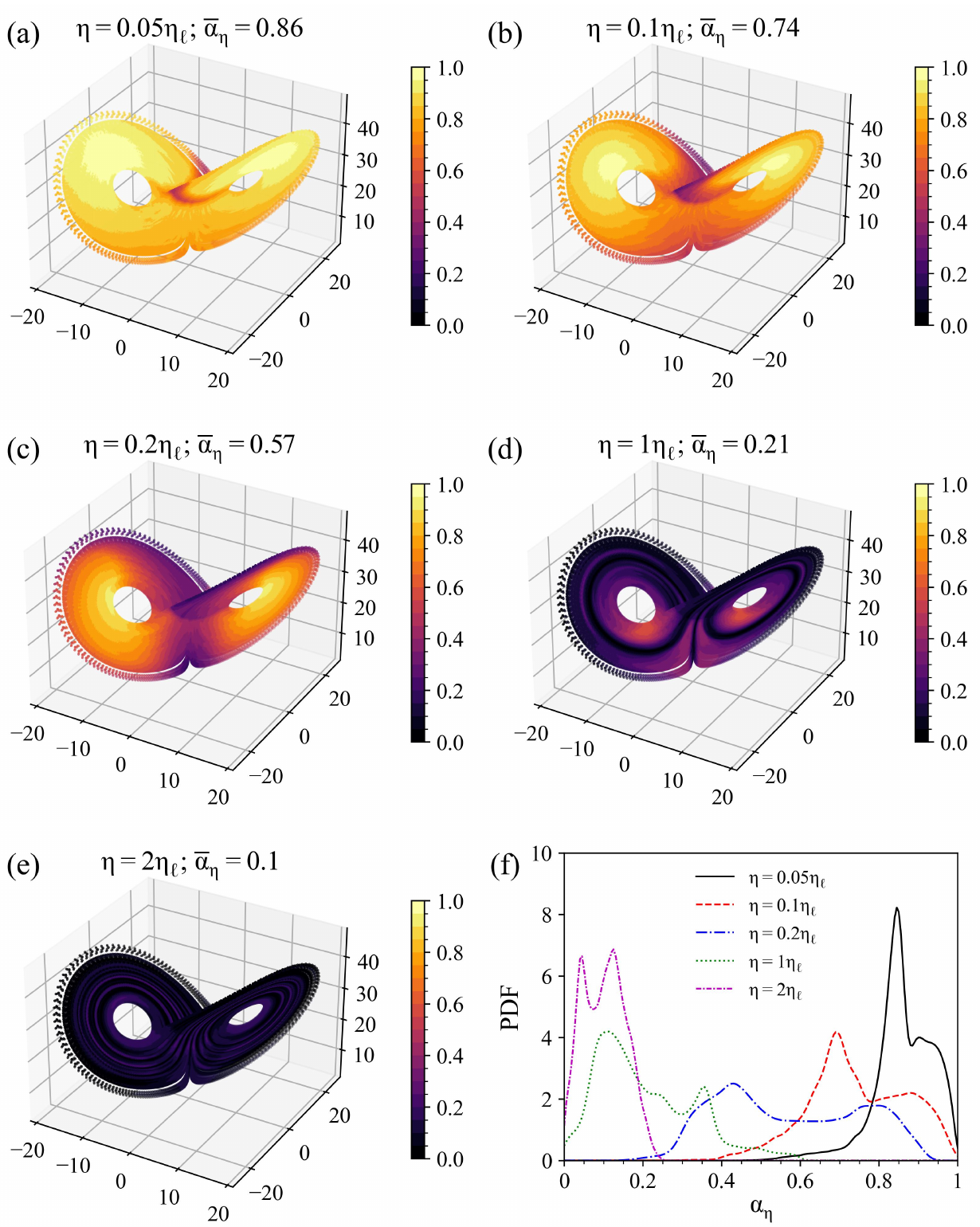}
\caption{Distribution of $\alpha_{\eta}$ at different prediction horizons for the Lorenz-63 system. (a-e) Lorenz attractor colored by $\alpha_{\eta}$ values at five different forcasting horizons: $\eta = [0.05, 0.1, 0.2, 1, 2]\eta_{\ell}$, each corresponding to timesteps: $L = [11, 22, 44, 220, 440]$. (f) Probability distribution of $\alpha_{\eta}$ at the same five different forecasting horizons. The quantile $q$ applied in this analysis is 0.99, and the Theiler window size $w$ is set to 50 time steps.}
\label{fig:fig2}
\end{figure}

We consider the Lorenz-63 attractor as an example of idealized dynamical system. 
The data for this test case is obtained from the numerical simulation of the Lorenz-63 system, with a time resolution of \num{5e-3} over \num{1e5} time steps, using the classical set of parameters that leads the system to be chaotic (see SI Appendix section 1 for details). 
Figure~\ref{fig:fig2} shows the distribution of $\alpha_{\eta}$ for the Lorenz-63 system, for five different forecasting horizons: $\eta = [0.05, 0.1, 0.2, 1, 2]\eta_{\ell}$.
The quantity $\eta_{\ell}$, that represents the Lyapunov time, is equal to $1.1$ time units and is computed as the inverse of the maximum Lyapunov exponent of the Lorenz-63 system.

Figs.~\ref{fig:fig2}(a-e) depict the Lorenz attractor in the phase space colored by the values of $\alpha_{\eta}$, for the 5 different forecasting horizons considered. 
For each panel in Fig.~\ref{fig:fig2}(a-e), the title on the top reports the forecasting horizon, and the average of the local predictability, namely $\bar{\alpha}_{\eta}$.  
We note how, as we increase the forecasting horizon, $\bar{\alpha}_{\eta}$ tends to decrease, as also depicted by the increasingly darker colors from Fig.~\ref{fig:fig2}(a) to Fig.~\ref{fig:fig2}(e). 
This result is in agreement with what we shall expect -- the longer the forecasting horizon, the less predictable the system. 
This conclusion is also summarized in Fig.~\ref{fig:fig2}(f), that shows the distribution of $\alpha_{\eta}$ for the five different forecasting horizons: the two longest horizons display a distribution peaked on low values of predictability ($\sim 0.2$), while shorter horizons tend to display distributions peaked toward higher values of predictability. 
To further test the resilience of $\alpha_{\eta}$ to noise, we applied $\alpha_{\eta}$ to the Lorenz-63 system with stochastic diffusion terms for comparison (see SI Appendix Fig. S1). The results are similar to Fig.~\ref{fig:fig2}, while we observe small-scale fluctuations of $\alpha_{\eta}$ over the attractor and a more rapid drop in average predictability, as one shall expect. 

These results confirm that the predictability of the Lorenz-63 system is closely related to the local reference state in the phase space and the forecasting horizon.
In addition, the phase-space distribution of $\alpha_{\eta}$ shows some interesting features, namely: for short forecasting horizons the region where the Lorenz-63 attractor switches lobes seem to have low predictability (Fig.~\ref{fig:fig2}(a)), while for longer forecasting horizons, also the wings of the attractor become less predictable (Fig.~\ref{fig:fig2}(b,c)). 
These findings are consistent with previous studies on local predictability of Lorenz systems~\cite{nese1989quantifying,palmer1993extended}.
Remarkably, for forecasting horizons twice the Lyapunov time, the overall system loses predictability (Fig.~\ref{fig:fig2}(e)).

\begin{figure}[h!]
\centering
\includegraphics[width=0.75\columnwidth]{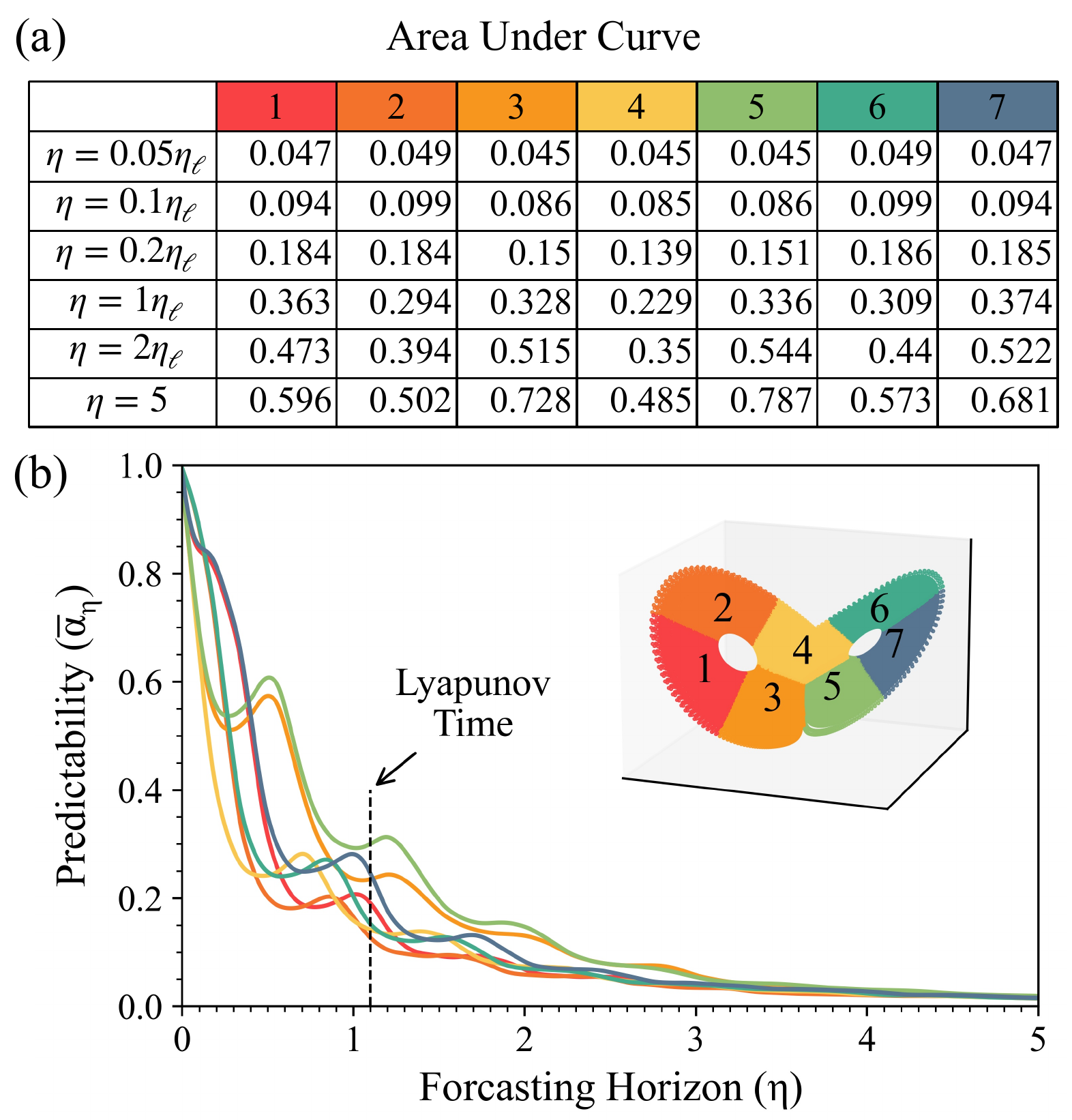}
\caption{Temporal and phase-spatial variations of $\alpha_{\eta}$. 
The evolution of $\alpha_{\eta}$ averaged across seven clusters of states on the Lorenz-63 attractor. 
Inset: Lorenz-63 attractor colored to represent seven clusters obtained from K-means algorithm, each labeled with corresponding numbers.}
\label{fig:fig3_clusters}
\end{figure}

To further show how $\alpha_{\eta}$ reveals phase-space and time features that can be used for understanding the predictability of dynamical systems, we present the evolution of the average $\alpha_\eta$ for seven different clusters (or regions) of the Lorenz-63 attractor. 
These clusters are depicted in Fig.~\ref{fig:fig3_clusters}, and are obtained by applying a K-means clustering method to the Lorenz-63 data used for Fig.~\ref{fig:fig2}, similar to the one adopted in~\cite{kaiser2014cluster}.
Interestingly, the clustering identifies portions of the attractor in the phase space that seem to match some of the observations made for Fig.~\ref{fig:fig2}.
In particular, cluster 4 represents the region where trajectories are switching lobes, clusters 1, 2, 6, and 7 represent the wings of the attractor, while clusters 3 and 5 represent intermediate regions that contain features from the five clusters just described. 
We note how, as expected, cluster 4 has the lowest short-term predictability among the regions considered, with a sudden drop at short forecasting horizons (yellow line).
The wings of the attractor follow, with clusters 2 (dark orange) and 6 (green) being the least predictable after cluster 4, immediately followed by clusters 1 (red) and 7 (blue). 
Clusters 3 and 5, after a sudden predictability drop similar to the one experienced for cluster 4, maintain relatively good predictability for larger forecasting horizons compared to all other clusters. 
This is confirmed, if we look at the area under each curve (AUC) reported in Fig.~\ref{fig:fig3_clusters}, that represent the average predictability of each cluster with respect to the forecasting horizons, with clusters 3 and 5 having the highest average predictability, and cluster 4 the lowest.  

We also note that, despite the general trend of predictability decreasing as the prediction horizon increases, $\alpha_{\eta}$ does not always decrease monotonically. 
There are instances showing returns of predictability, as also reported in~\cite{smith1999uncertainty}, and as reflected more clearly in Fig.~\ref{fig:fig3_clusters}(b); see also discussion above regarding the relationship between the definitions provided in Eqs.~\ref{eq:conditional} and~\ref{Lyapunovfinite}.

We additionally compare our results with the nonlinear local Lyapunov exponent (NLLE)~\cite{ding2007nonlinear,huai2017quantifying}, an established index used to quantify local predictability that we find closely relevant, as reported in SI Appendix section 4. 
The results for the NLLE display some notable differences with our local predictability index while also showing some similarities (see SI Appendix Fig. S6). 
These differences are arguably due to the use of different perspectives in defining predictability. Their framework stems from the Lyapunov exponent, focusing on the exponential error growth rate, while our method is rooted in the concept of \textit{recurrences} in dynamical system theory. We acknowledge that both methods have their own caveats: NLLE predefines the rule of error growth, while our method neglects the information on trajectories that have left the neighborhood. To overcome this, we propose an extension of our predictability index by accounting for the information given by the distance of all the neighboring trajectories -- i.e., both those that remained close to the reference state and those that departed from it, as shown in SI Appendix section 5.

\begin{figure}[h!]
\centering
\includegraphics[width=0.85\columnwidth]{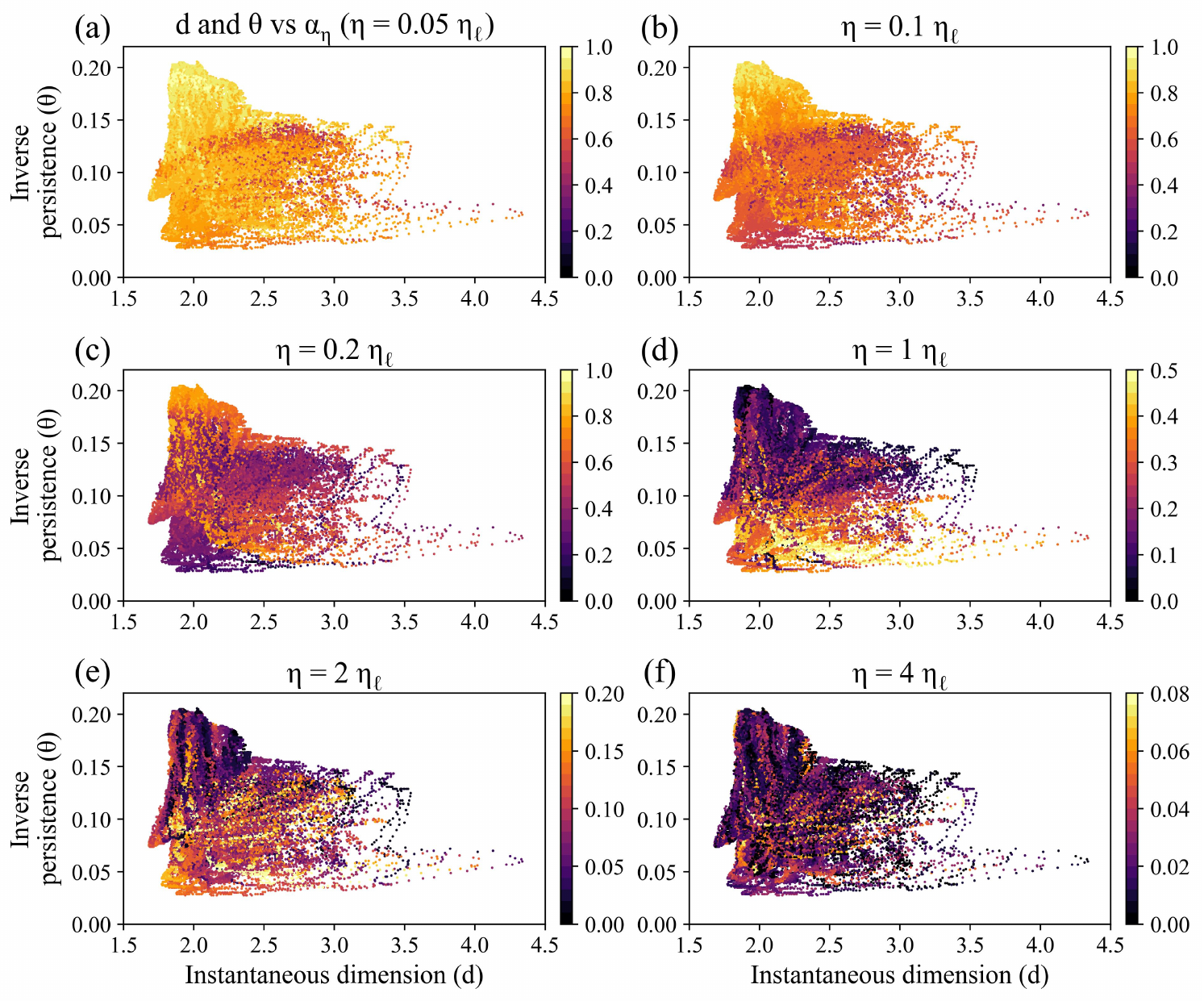}
\caption{Comparative analysis of $\alpha_{\eta}$ with local dynamical indices for the Lorenz-63 system. (a-f) Scatter plots of local dimension versus local inverse persistence, where each dot represents one state. All states are colored based on $\alpha_{\eta}$ values for $\eta = [0.05, 0.1, 0.2, 1, 2, 4]\eta_{\ell}$. Note that the color scale is different in the last three panels in order to improve readability.}
\label{fig:fig4_lorenz_scatter}
\end{figure}

Furthermore, we evaluated the relationship of the previously introduced local dynamical indices, $d$ and $\Theta$, to predictability, by comparing them against $\alpha_\eta$. 
In Fig.~\ref{fig:fig4_lorenz_scatter}, we show scatter plots of the local predictability index $\alpha_{\eta}$ versus the local dimension $d$ and the inverse persistence $\theta$. 
We note that the intuition that high dimension and low persistence means low predictability does not hold for the case of the Lorenz-63 system.
Specifically, states on the two wings of the Lorenz attractor exhibit low persistence (high values of $\theta$), indicating that these states are likely to leave the neighborhood immediately~\cite{faranda2017dynamical}. However, they also demonstrate high predictability over short forecasting horizons. This is not surprising when considering the dynamics on the wings of the Lorenz attractor—although these states may leave the neighborhood quickly, their neighboring trajectories do not diverge significantly, making them predictable in short forecasting horizons. 

We show, in the next section, that for the real-world data considered, the agreement between the interpretation of the $d$-$\theta$ space and $\alpha_\eta$ seems to hold instead.

\subsection*{Atmospheric Circulation}\label{subsec:real-world}

In addition to the test case presented for idealized systems, we computed $\alpha_{\eta}$ using real-world data. In particular, we focused on studying the predictability of large-scale atmospheric circulation patterns in the Euro-Atlantic sector (see Materials and Methods for details). 
In this context, we consider the Z500 map for each day as a state along the system’s trajectory in the approximated phase space. We emphasize that this region serves as an ideal testbed to validate our predictability index due to its extensive body of relevant research, especially considering that no state-dependent predictability index is directly comparable for high-dimensional datasets.

In the Euro-Atlantic sector, the definition of weather regimes offers a direct way of classifying atmospheric states.
Weather regimes are defined as recurrent and quasi-stationary large-scale circulation patterns~\cite{michelangeli1995weather,Springer2024}. 
Despite being a coarse characterization of atmospheric variability, weather regimes are found to be associated with conditions that are specifically favorable for certain types of weather extremes~\cite{yiou2004extreme}, and they have exhibited varying levels of predictability in real-time forecasting~\cite{ferranti2015flow}. 
In this work, we have adopted the four-regime definition widely used for scientific research~\cite{vautard1990multiple}, although various definitions for weather regimes exist~\cite{vautard1990multiple, grams2017balancing}.
Based on this definition, all extended wintertime (DJFM) Z500 daily patterns are classified into five different categories including NAO+, Alantic Ridge (AR), NAO-, Scandinavian Blocking (SB) and no regime, using a method slightly adapted from~\cite{cassou2008intraseasonal} (see SI Appendix section 6 for details). 
\begin{figure}[h!]
\centering
\includegraphics[width=0.75\columnwidth]{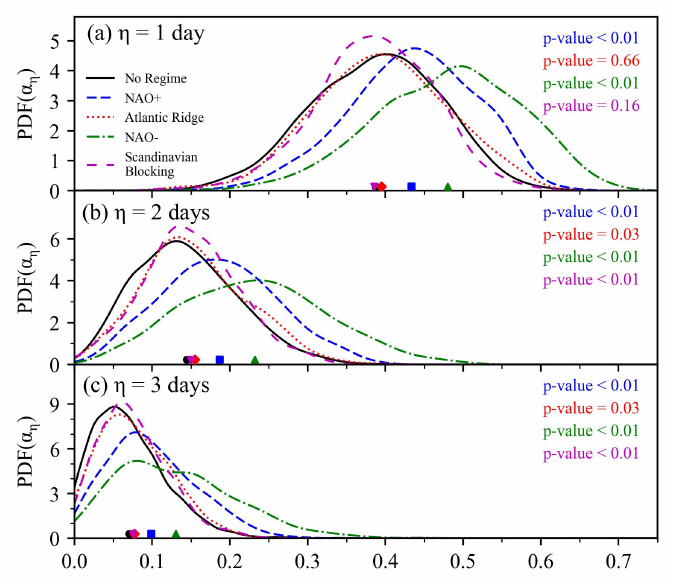}
\caption{Predictability analysis of North Atlantic wintertime weather regimes. (a-c) Probability distribution of $\alpha_{\eta}$ for four weather regime and no regime: NAO+ (blue dashed line), Atlantic Ridge (red dotted line), NAO- (green dot-dashed line), Scandinavian Blocking (magenta loosely dashed line), and no regime (black line), for $\eta=1-3$ days. The markers on the x-axis show the mean value of $\alpha_{\eta}$ for different categories. The statistical significance is evaluated between the distribution of each weather regime and that of no regime using the Kolmogorov-Smirnov test. The quantile $q$ applied in this analysis is 0.99, and the Theiler window size $w$ is set to 7 days.}
\label{fig:fig5_WR}
\end{figure}
Figure~\ref{fig:fig5_WR} shows the distribution of $\alpha_\eta$ across all five categories for $\eta=1-3$ days. 
The results indicate that the NAO- regime has the highest predictability, followed by NAO+, with the remaining two regimes showing no significant difference from having no regime for $\eta=1$ day.
For $\eta=2$ and $3$ days, NAO+ and NAO- still maintain notably higher mean values compared to other categories, with the no regime category having the lowest values, confirming that large scale weather regimes are a primary source of predictability in the Euro-Atlantic region. 
These results are in excellent agreement with previous investigation dedicated to the Euro-Atlantic sector: NAO- exhibits stronger predictability than other regimes, while the blocked zonal regimes (SB and AR) are notoriously unpredictable~\cite{ferranti2015flow,faranda2017dynamical,ECMWF2018}. 
Additionally, we find further confirmation of the general fact that blockings are associated with anomalously high instability compared to zonal flow conditions~\cite{Schubert2016,LucariniGritsun2020}.
These findings further validate the effectiveness of our index, supported by both the underlying physical explanations and the model-based prediction results. 
We note that the analysis just presented can be extended to any other variables and domains, with a key region of interest being the tropical Indo-Pacific, given the complex interaction of different modes of variability~\cite{dong2024indo}.
\begin{figure}[h!]
\centering
\includegraphics[width=0.9\columnwidth]{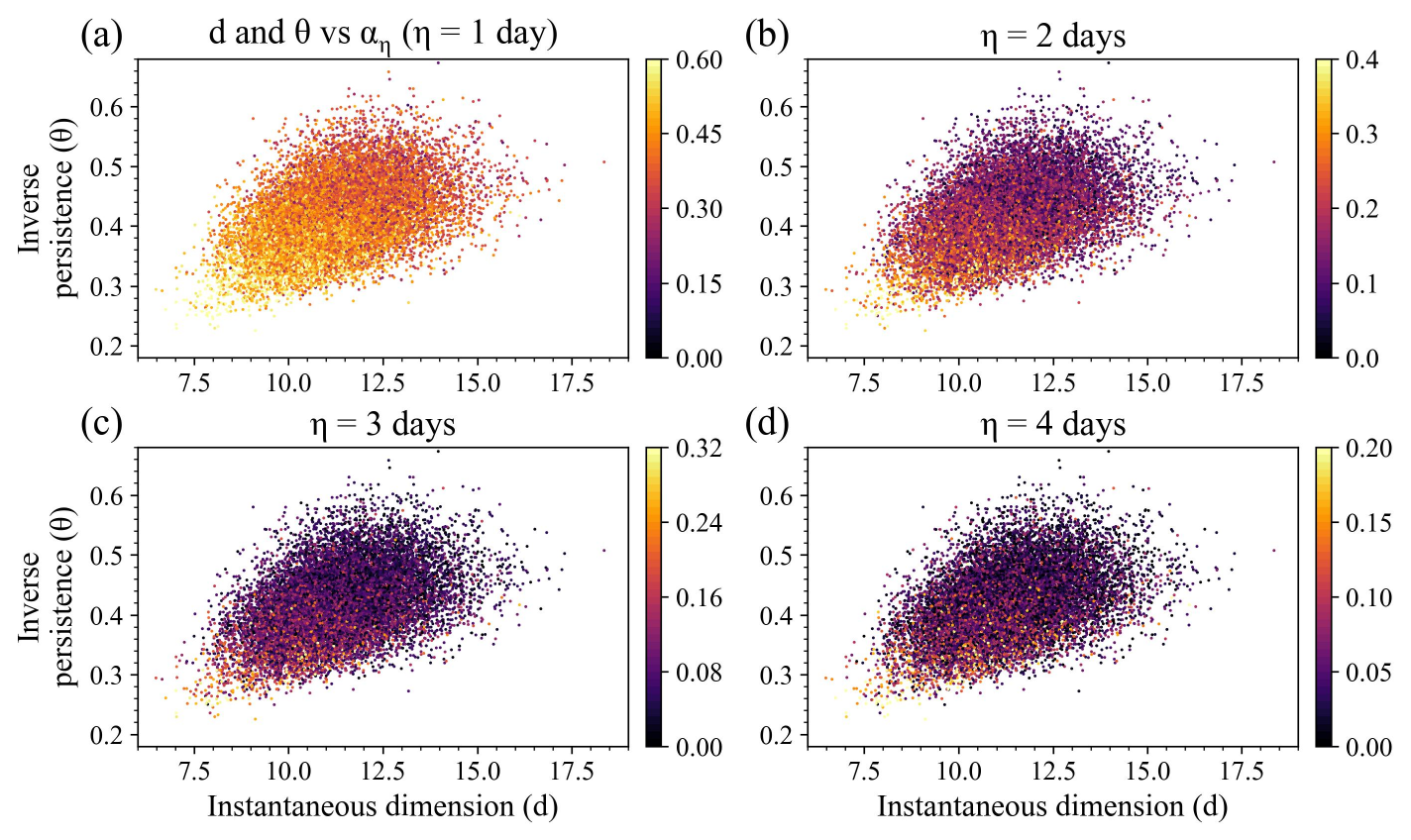}
\caption{Comparative analysis of $\alpha_{\eta}$ with local dynamical indices for Z500 patterns in the Euro-Atlantic sector. (a-d) Scatter plots of local dimension versus local inverse persistence, where each dot represents one daily Z500 pattern. All states are colored based on $\alpha_{\eta}$ values for $\eta=1-4$ days. Note that the color scale is different in each panel in order to improve readability.}
\label{fig:fig6_d_theta} 
\end{figure}

Fig.~\ref{fig:fig6_d_theta} illustrates scatter plots of local dimension $d$ vs inverse persistence $\theta$ for all the atmospheric states, colored by the predictability index $\alpha_{\eta}$, for different prediction horizons ($\eta$) ranging from 1 to 4 days. 
The overall predictability drops as $\eta$ increases, as expected due to the highly chaotic nature of atmospheric circulation.
Notably, states in the lower-left corner of the $d$-$\theta$ scatter plot space consistently exhibit higher-than-average predictability, regardless of the value of $\eta$. 
This aligns with the assumption in previous works for which states with low local dimension and high persistence are most predictable. 
From the dynamical system theory perspective, a lower local dimension indicates a simpler dynamics, while higher persistence means that a state is likely to remain similar to its current configuration for a longer time.
This result confirms our previous hypothesis while also further demonstrating the effectiveness of the proposed predictability index for high-dimensional systems.

\subsection*{Other systems}\label{subsec:other-systems}

\begin{figure}[h!]
\centering
\includegraphics[width=0.85\columnwidth]{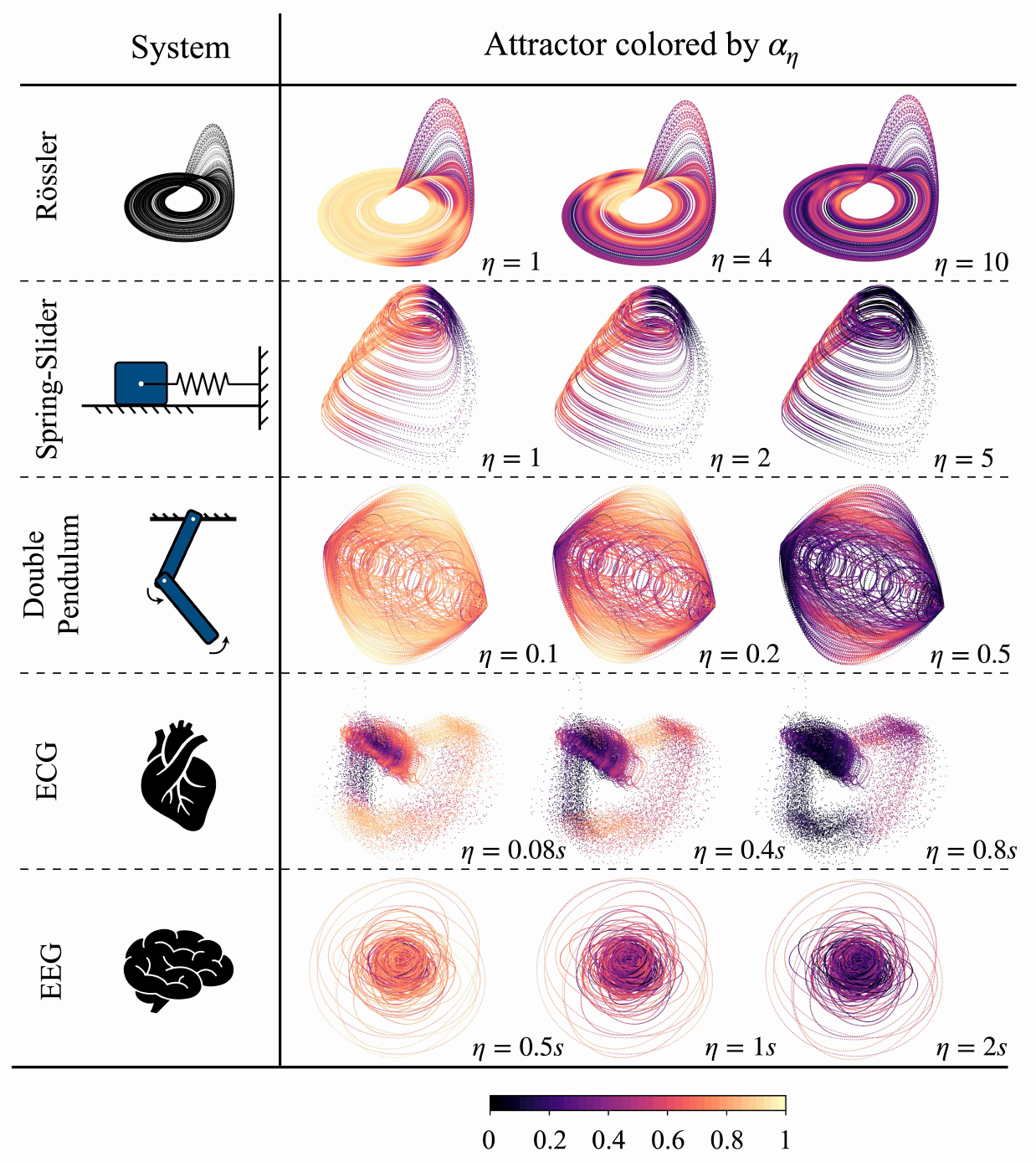}
\caption{Summary of $\alpha_\eta$ applied to various systems, including another canonical dynamical system (the Rössler attractor), simulated slow earthquakes (spring-slider model), a classical physics problem (the double pendulum), and real-world biological data (observational ECG and EEG data). The forecasting horizon $\eta$ is annotated at the bottom-right corner of each attractor. All the attractors presented here are three dimensional for demonstration purposes, but their corresponding $\alpha_\tau$ are computed in higher-dimensional spaces, as detailed in SI Appendix section 7. The quantile $q$ applied in all analysis is 0.99, and the Theiler window size $w$ is set to 50 time steps.}
\label{fig:other_systems}
\end{figure}

In addition to the Lorenz-63 system and climate data, in Fig.~\ref{fig:other_systems}, we demonstrate the applicability of $\alpha_\eta$ across a diverse range of systems. These include another canonical dynamical system (Rössler attractor), simulation data for slow earthquakes (spring-slider), a classical physics problem (double pendulum), and real-world biological data (observational ECG and EEG data). The stochastic spring-slider system was originally introduced to simulate slow earthquakes, which feature a random attractor governed by stochastic differential equations, as detailed in~\cite{gualandi2023deterministic}. The final three examples use the same dataset as~\cite{brunton2017chaos}, with the double pendulum data generated through numerical simulation, while the other two are obtained from observational data.

The results for the Rössler attractor show a certain degree of similarity to those of the Lorenz-63 system: the Lorenz-63 system experiences a significant drop in predictability before lobe transitions, whereas the Rössler attractor exhibits a similar phenomenon prior to bursting events in the z-direction. 
We also note that the region following the bursting events exhibits markedly higher predictability, as the trajectories passing through this region will oscillate predictably in the x–y plane until re-entering the region prior to the bursting events. In the case of the spring-slider system, there is high predictability during the almost linear inter-seismic period. The time-lagged recurrence index starts decreasing when the slip rate of the slider increases. The fact that the onset of the rupture is marked by high unpredictability is in line with the idea that it is a difficult task to predict the final size of the event at the onset of the slipping phase. For the attractor of the double pendulum, we note that two endpoints of the attractor show noticeably lower predictability, whereas the transitional regions between them show relatively higher predictability, particularly for the outer parts of the attractor. In the two real-world biological data examples, we similarly observed that $\alpha_\eta$ reveals variability in predictability within the phase space across different forecasting horizons. These insights can improve the analysis of such systems and deepen our understanding of their underlying dynamics. The applications to these diverse systems collectively demonstrate that this novel method has broad applicability and effectiveness. 

\subsection*{A note on real-time predictability}\label{subsec:real-time}

Our findings demonstrate the agreement of $\alpha_{\eta}$ with known results, positioning it as a viable index for analyzing the local predictability of dynamical systems. 
Yet, one key point remains outstanding: $\alpha_{\eta}$, as is, cannot be used for real-time applications, as it requires the future state of the system. 
However, since $\alpha_\eta$ quantifies local predictability information in the phase space, for a given unprecedented state $\boldsymbol{\zeta}^{\prime}$, we can compute the average $\alpha_\eta$ of its analogues (i.e., similar states to $\boldsymbol{\zeta}^{\prime}$ that appeared in the past). 
The number of analogues can be set once again using a quantile threshold, but this could be different from the one used to define the local properties.
In practice, if we consider $N_a$ analogues, indicated by $\lbrace {\bf{x}}(t_{a_i}) \rbrace_{i=1}^{N_a}$, with $t_{a_i}$ being the time at which the analogue number $a_i$ occurred, the predictability index proxy is:
\begin{equation}
\hat{\alpha}_{\eta}(\boldsymbol{\zeta}^{\prime}) = \frac{1}{N_a} \sum_{i=1}^{N_a}{\alpha_{\eta}({\bf{x}}(t_{a_i}))}
\label{eq:alphat-proxy}
\end{equation}
This can be used for real-time predictability analyses of dynamical systems. 
For instance, in the context of weather and climate applications, we can identify the analogues of today's data, and compute the predictability index proxy $\hat{\alpha}_{\eta}(\boldsymbol{\zeta}^{\prime})$ using Eq.~\ref{eq:alphat-proxy}.
This will give us the real-time predictability of the system as of today, thereby allowing us to adopt the index for real-time applications. 
An example of real-time applications is presented in SI Appendix section 8, where we show an application for the Lorenz-63 system, with the real-time predictability proxy $\hat{\alpha}_{\eta}(\boldsymbol{\zeta}^{\prime})$ successfully capturing the phase-space and time features of the predictability.

\section*{Conclusions}
\label{sec:conclusions}

This work complements the rich body of work on the pivotal topic of predictability of dynamical systems, from the pioneering works of Poincaré and Lorenz, to modern dynamical system theory and its applications to real-life complex systems, especially in the context of weather and climate prediction.
In particular, we provide a time-lagged recurrence index that can infer state-dependent predictability from high-dimensional observational datasets, overcoming the curse of dimensionality that affects more traditional approaches. While our approach has a resemblances with classical data-driven methods used for performing Lyapunov analysis on general systems \cite{Wolf1985,Rosenstein1993}, the local predictability index is indeed novel and is able to describe flexibly key dynamical features of complex systems.
We applied this index to dynamical systems of varying complexity, ranging from idealized low-dimensional systems to real-world high-dimensional atmospheric reanalysis datasets and 
we showcased its effectiveness. In particular, our findings confirm that when in the Euro-Atlantic sector of the mid-latitudes a blocking is present, the predictability of the atmosphere is typically anomalously low. 
We further showed how the new index reveals the scale-dependent nature of predictability by using different quantiles, and how it can be used in real-time problems. 
Our approach also addresses the ambiguity in terms of predictability that surrounds the two local dynamical indices, namely dimension $d$ and persistence $\Theta$, providing a framework that directly addresses predictability. 

We conclude this work with a note. 
The definition of $\alpha_\eta$, does not require $\eta > 0$.
Therefore, $\alpha_\eta$ can be computed backward in time using $\eta < 0$.
In this case, $\alpha_\eta$ characterizes the dynamics of the system as it approaches the reference state of interest, and it provides information that are entirely different from the forward-in-time analysis presented so far, using $\eta > 0$.
Indeed, the backward predictability analysis can be used for understanding the dynamical pathways leading to extreme events as well as other phenomena in dynamical systems~\cite{li2023backward, noyelle2024investigating}.
As a methodological study, we consider such in-depth analysis to be beyond the scope of this work, but we believe it may shed new light on various scientific domains.

\section*{Materials and Methods}

\subsection*{Local dynamical indices}\label{subsec:dynamical-indices}

In the framework proposed by Lucarini et al.~\cite{lucarini2016extremes}, two dynamical indices, the local dimension $d$ and persistence $\Theta$, were introduced to characterize the local properties of dynamical systems.
To compute these indices, we first consider the negative logarithmic returns $g\left({\boldsymbol{\zeta}}, {\bf{x}}(t)\right)$, as defined in Eq.~\ref{eq:log_dist}, which quantifies proximity to a given state $\boldsymbol{\zeta}$. 
The exceedance ${\bf{u}}({\boldsymbol{\zeta}})$, introduced in Eq.~\ref{eq:exceedance}, is defined for the neighboring states (i.e., recurrences) of $\boldsymbol{\zeta}$ and measures how much $g\left({\boldsymbol{\zeta}}, {\bf{x}}(t)\right)$ exceeds a high quantile $q$.
According to extreme value theory~\cite{freitas2010hitting,lucarini2012universal}, assuming the independence of the exceedances, the cumulative probability distribution $F({\bf{u}}, {\boldsymbol{\zeta}})$ follow the exponential member of the generalized Pareto distribution (GPD), that is:
\begin{equation}\label{eq:cumulative}
F({\bf{u}}, {\boldsymbol{\zeta}}) \sim \exp\left[-\frac{{\bf{u}}({\boldsymbol{\zeta}})}{\sigma({\boldsymbol{\zeta}})}\right].
\end{equation}
The parameter $\sigma({\boldsymbol{\zeta}})$, which is the scale parameter of the distribution, depends on the state $\zeta$ in phase space and can be used to derive the local dimension via $d({\boldsymbol{\zeta}}) = 1/\sigma({\boldsymbol{\zeta}})$~\cite{lucarini2016extremes}. In practice, the local dimension is computed by explicitly fitting an exponential distribution to the exceedances $\bf{u}({\boldsymbol{\zeta}})$, see discussion in \cite{datseris2023estimating} and \cite{pons2023statistical}.
The estimate of the attractor's dimension derived from $d({\boldsymbol{\zeta}})$ seems to be more efficient than previous methods based on evaluating correlation integrals~\cite{Grassberger1983,Ding1993}.
The other local index, persistence ($\Theta$), is defined as the inverse of the extremal index $0 \leq \theta \leq 1$~\cite{freitas2012extremal,moloney2019overview}, a dimensionless parameter that measures the inverse of the length of clustering of extremes. In this context, extremes are defined as the \textit{recurrences} within the neighborhood around a given state in phase space. Therefore, persistence ($\Theta$) estimates the average number of consecutive recurrences, namely the time steps for the system to resemble a given state in its neighboring phase space.

This framework is the basis for the new predictability index proposed in this work, and introduced in the main text above.

\subsection*{500hPa Geopotential Data in the Euro-Atlantic Sector}\label{subsec:data} 
In this study, the atmospheric circulation dynamics in the Euro-Atlantic sector (80°W to 50°E, 22.5°N to 70°N) served as a testbed to evaluate the applicability of $\alpha_\eta$ on a real-world high-dimensional dataset. 
We used the state-of-the-art ERA5 reanalysis data~\cite{hersbach2020era5} from 1979 to 2022, with a spatial resolution of 0.25°. 
The daily mean 500hPa geopotential height (Z500) was selected as the representative variable, which has been frequently used to characterize atmospheric circulation dynamics in the mid-latitudes~\cite{Benzi1986,MoGhil1988,DellAquila2005,jezequel2018role,Springer2024}. 
For this spatio-temporal system, we treat each daily spatial map of Z500 as an individual state, and the corresponding space can be regarded as an approximation to the phase space. 
This approach is widely adopted in practical applications involving real-world complex systems because accessing all variables of a system to construct its full phase space is often impossible.
Instead, we can focus on a subset of available observables to approximate the system's dynamics.
In this way, the evolution of atmospheric circulation patterns can be viewed as a trajectory within this high-dimensional space, thereby enabling the application of $\alpha_\eta$ to assess its state-dependent predictability.

\subsection*{Data and Software Availability}\label{subsec:availability}
The code for computing the local predictability index and the data used in this study are available at \url{https://github.com/MathEXLab/TLR}. The ERA5 reanalysis data used in this study are available at \url{https://cds.climate.copernicus.eu/#!/search?text=ERA5&type=dataset.} The data used for the double pendulum, ECG, and EEG are from~\cite{brunton2017chaos}.

\section*{Acknowledgments}
G.M. acknowledges support from MOE Tier 2 grant no 22-5191-A0001-0, titled: `Prediction-to-Mitigation with Digital Twins of the Earth’s Weather'. V.L. acknowledges the support provided by the Horizon Europe Project ClimTIP (Grant No. 100018693) and by the EPSRC project LINK (Grant No. EP/Y026675/1).

\bibliographystyle{unsrt}  
\bibliography{references}  

\clearpage
\section*{Supporting Information Appendix}
\appendix
\renewcommand{\thefigure}{S\arabic{figure}}
\setcounter{figure}{0}

\section{Lorenz system and its stochastic version}\label{sec:lorenz63}

The Lorenz-63 system~\cite{lorenz1963deterministic} was originally derived from a simplified model of the atmospheric convection process and has been widely used as an example to study chaotic dynamical systems. Its governing equations can be written as:
\begin{eqnarray}
\frac{dx}{dt} &=& \sigma(y-x), \\
\frac{dy}{dt} &=& x(\rho-z) - y, \\
\frac{dz}{dt} &=& xy - \beta z.
\end{eqnarray}
In this study, we adopt the classical parameters of the Lorenz system that exhibit deterministic chaos, where $(\sigma, \rho, \beta)=(10,28,8/3)$. Observational data are obtained through numerical simulation with a temporal resolution of $dt = 0.005$ over $10^5$ time steps. 

\begin{figure}[h!]
\centering
\includegraphics[width=0.8\columnwidth]{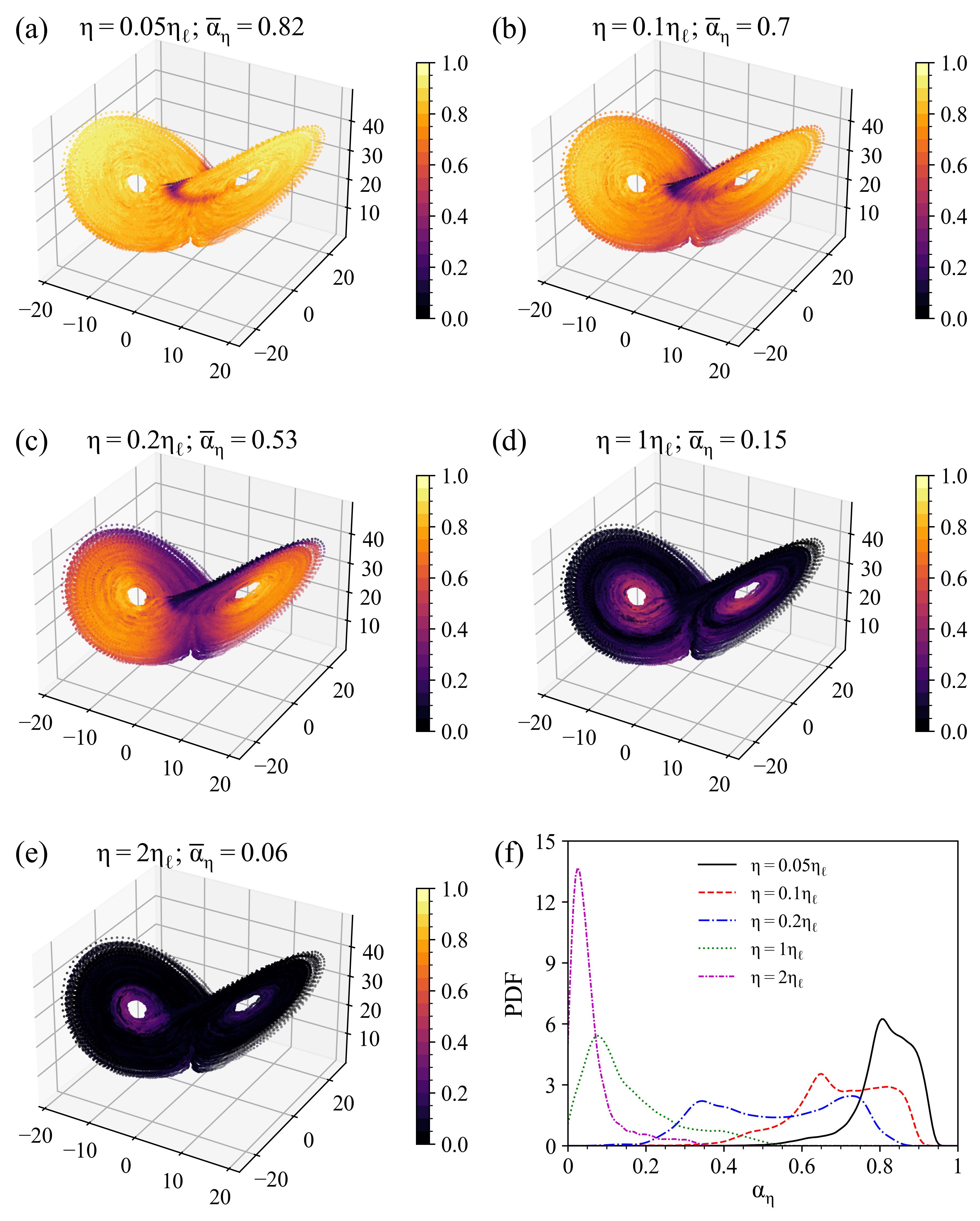}
\caption{Distribution of $\alpha_{\eta}$ at different prediction horizons for the stochastic Lorenz-63 system. Same as Figure 2, but for the stochastic version of the Lorenz-63 system. The quantile $q$ applied in this analysis is 0.99, and the Theiler window size $w$ is set to 50 time steps.}
\label{fig:lorenz_stochastic}
\end{figure}

Additionally, we test the impact of noise on our predictability index by considering the Lorenz-63 system with diffusion terms. The corresponding equations can be written as:
\begin{eqnarray}
\frac{dx}{dt} &=& \sigma(y-x) + \eta dW^1_t, \\
\frac{dy}{dt} &=& x(\rho-z) - y + \eta dW^2_t, \\
\frac{dz}{dt} &=& xy - \beta z + \eta dW^3_t.
\end{eqnarray}
where $dW^j_t$, $j=1,2,3$ indicate increments of three independent Wiener processes, and  $\eta$ modulates the intensity of the stochastic perturbations. In Fig.~\ref{fig:lorenz_stochastic}, we show the predictability index for the Lorenz system with stochastic perturbations, using the same numerical simulation settings as in Fig. 2, with the value of $\eta=1$. We note that the Fig.~\ref{fig:lorenz_stochastic} is similar to Figure 2 for the five different forecasting horizons considered. However, we observe small-scale fluctuations of $\alpha_{\eta}$, which corresponds to the characteristics of the stochasticity within the system. In addition, the value of $\alpha_{\eta}$ drops more rapidly on average as expected, since the stochastic nature will make the system less predictable in general.

\section{The scale-dependent nature of predictability}\label{sec:scale}

Just like other local indices, the computation of $\alpha_\eta$ requires a quantile parameter $q$ that defines \textit{recurrences}, indicating the number of \textit{recurrences} for each state used in the analysis. For local dimension $d$ and inverse persistence $\theta$, they are found to be stable within a certain range where the threshold is neither too high, making \textit{recurrences} insufficient to fit the generalized Pareto distribution, nor too low, leading to other distributions~\cite{faranda2017dynamical}. For $\alpha_\eta$, although we do not need to fit the generalized Pareto distribution, we are still interested in how the value of $q$ will affect the results. 

Different values of $q$ will result in different sizes of the neighborhood. Therefore, We interpret the predictability index derived from different $q$ actually representing local predictability at different scales. The scale-dependent nature of predictability has been widely recognized in previous studies~\cite{boffetta1998extension,aurell1997predictability,ding2007nonlinear}. For instance, in the linear regime, the maximum Lyapunov exponent denotes predictability for dynamical systems. However, when the error is finite (outside the linear regime), this is no longer sufficient because nonlinear dynamics play a more important role in predictability.

\begin{figure}[h!]
\centering
\includegraphics[width=0.8\columnwidth]{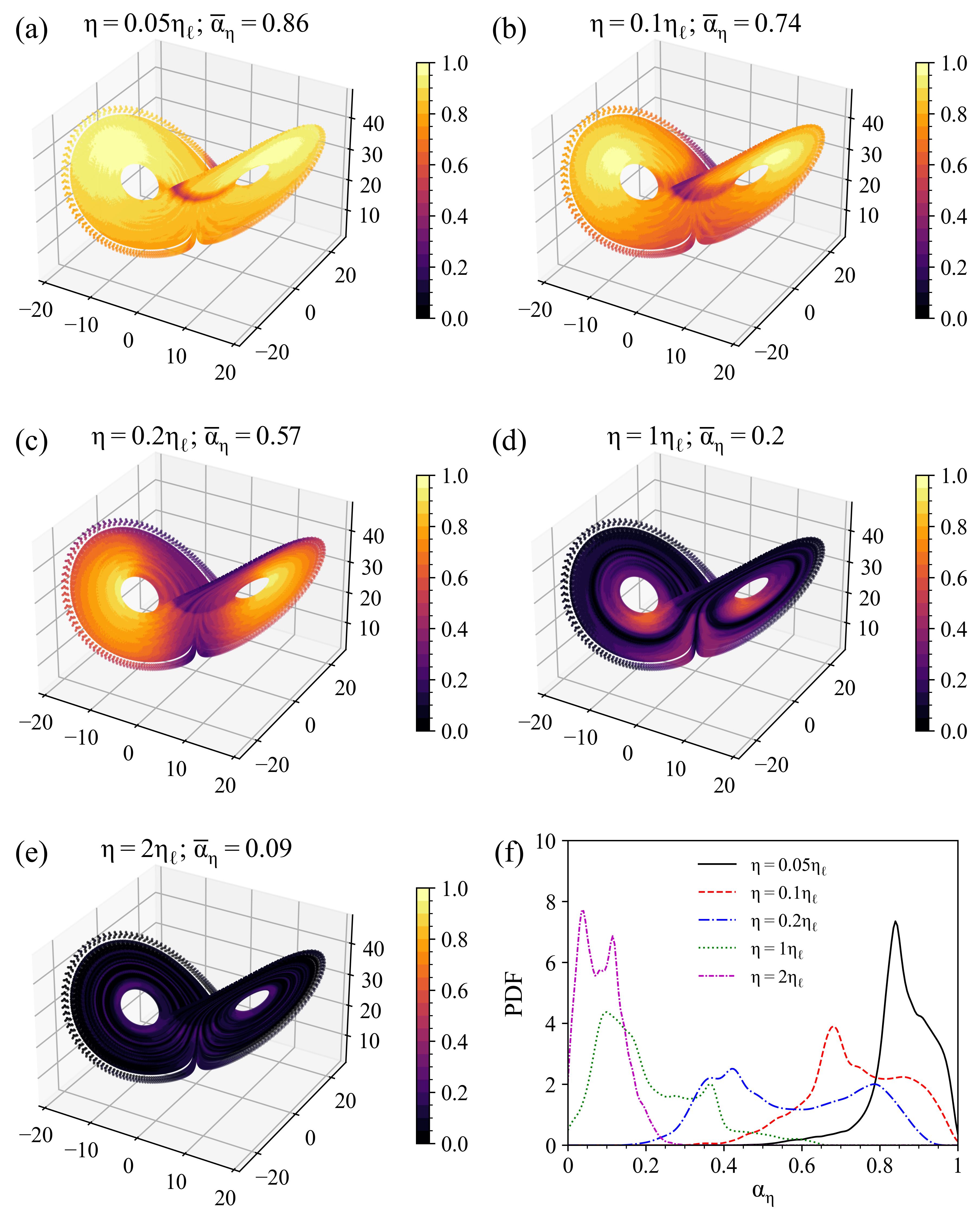}
\caption{Comparative analysis of the impact of different quantile choices ($q$) on $\alpha_{t}$. Same as Fig. 2, but for $q=0.98$. }
\label{fig:supfig_ql098} 
\end{figure}

\begin{figure}[h!]
\centering
\includegraphics[width=0.8\columnwidth]{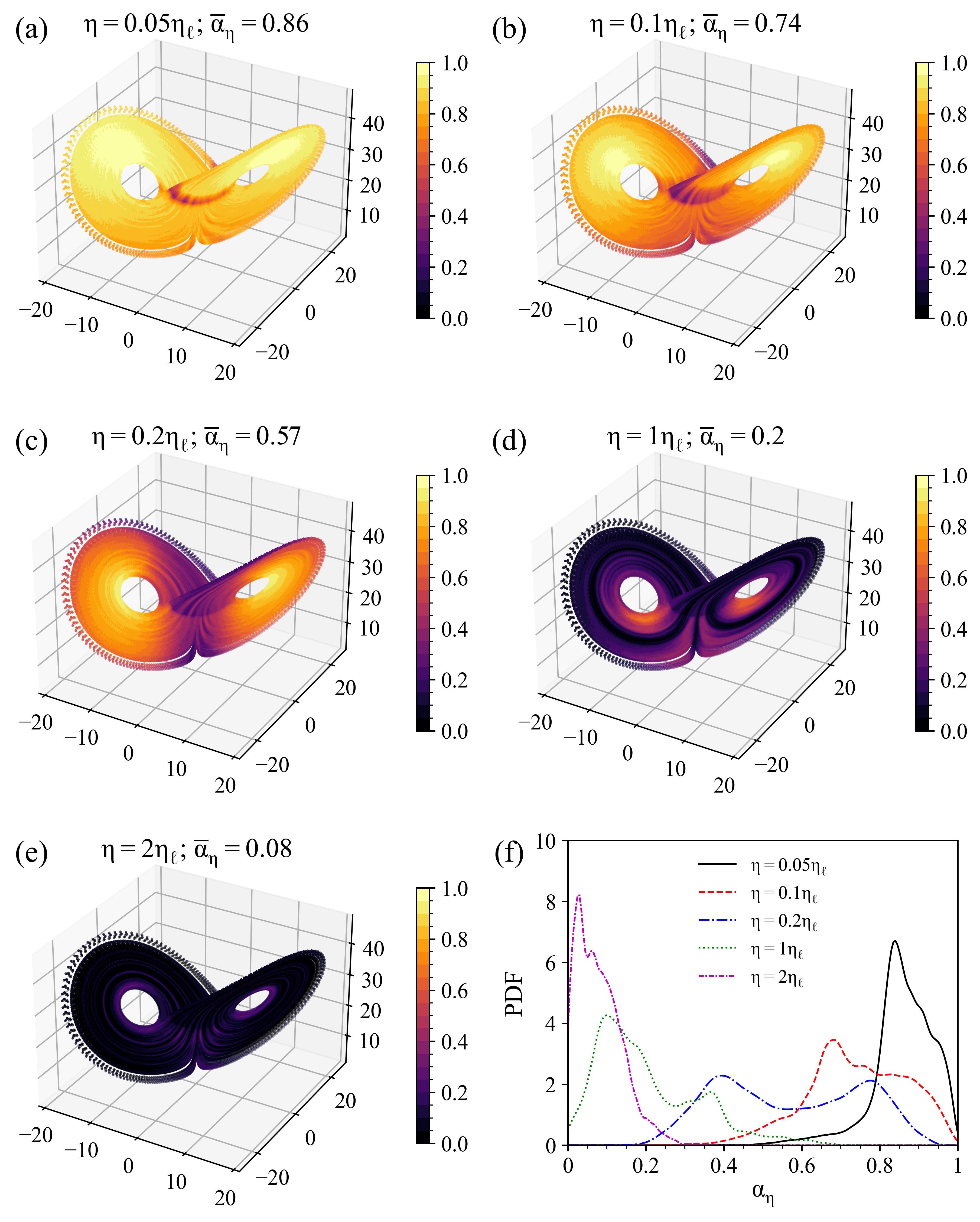}
\caption{Comparative analysis of the impact of different quantile choices ($q$) on $\alpha_{t}$. Same as Fig. 2, but for $q=0.995$.}
\label{fig:supfig_ql0995}
\end{figure}

In this paper, all the results displayed in the main text use $q$ = 0.99, which means the closest one percent of states to the reference state are selected as \textit{recurrences}. Figure~\ref{fig:supfig_ql098} and~\ref{fig:supfig_ql0995} show the results for $q=0.98$ and $q=0.995$, respectively, following exactly the same settings as Figure 2. We note that using different $q$ values yields similar overall results, with the most prominent features being noticeable in each case. However, we also observe some differences: for larger q values ($q=0.995$)
, small-scale stripes on the attractor are visible, whereas for relatively smaller q values, no such patterns exist and the variation of $\alpha_\eta$ in phase space appears smoother. This confirms the close relationship between predictability and scale, indicating that the choice of $q$ should be made according to specific application needs.

\begin{figure}[h!]
\centering
\includegraphics[width=1\columnwidth]{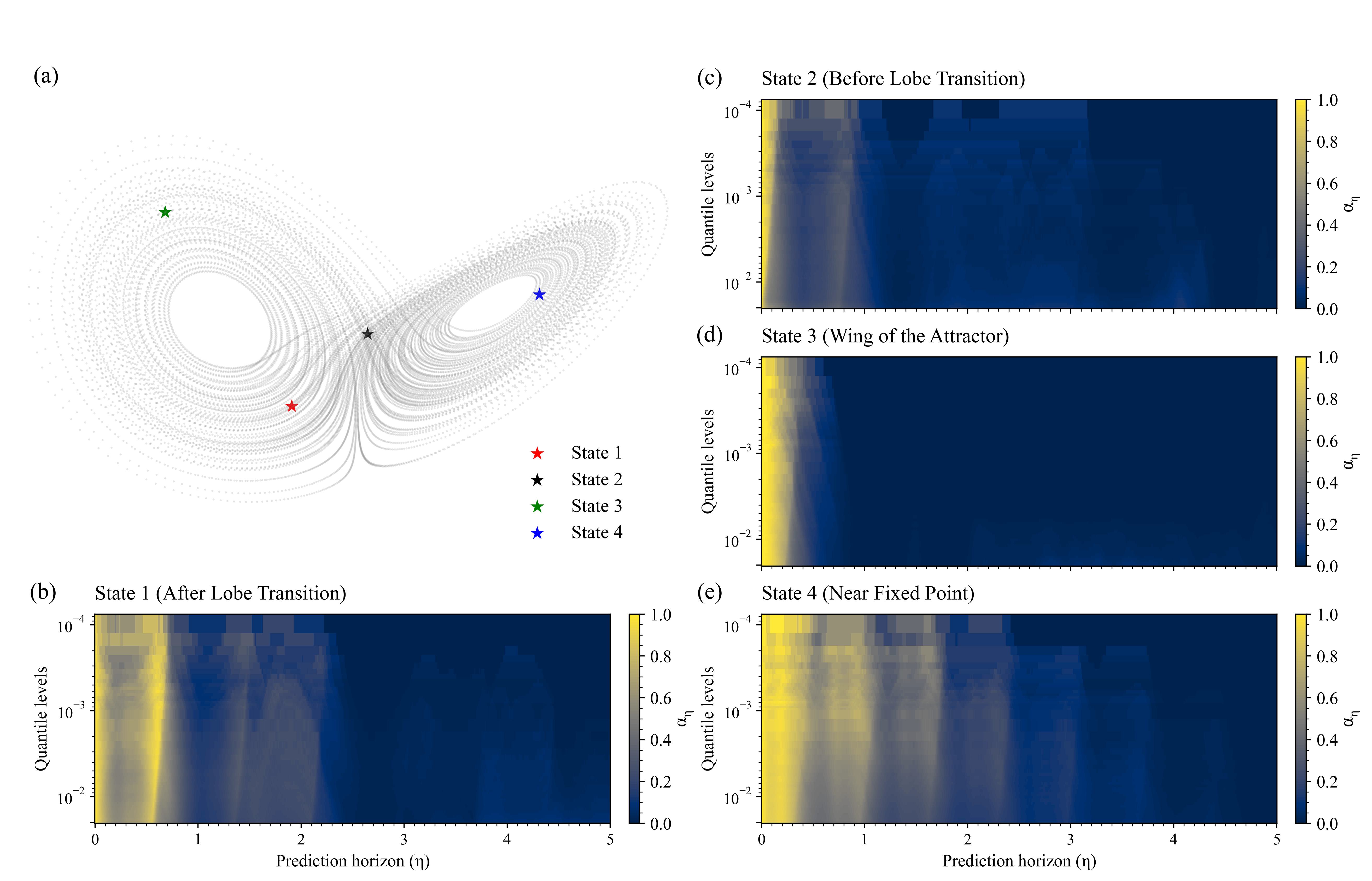}
\caption{Predictability analysis for different $q$ values. (a) Illustration of the locations of the states considered in the Lorenz-63 phase space. (b-e) Heatmaps of $\alpha_\eta$ for four states located at the wing, near the fixed point, before lobe transition (before the intersection of the two wings), and after lobe transition (after the intersection of the two wings), respectively. }
\label{fig:supfig_scale}
\end{figure}

In Figure~\ref{fig:supfig_scale}, we further tested a wider range of $q$ values for their influence on the value of $\alpha_\eta$. 
Four states at different locations on the Lorenz-63 attractor are shown in  Fig.~\ref{fig:supfig_scale} (a), representing the wing, near the fixed point, before lobe transition (before the intersection of the two wings), and after lobe transition (after the intersection of the two wings). 
The $\alpha_\eta$ computed from $q$ ranging from 0.98 (2000 $recurrences$ for dataset with length of $10^5$) to 0.9999 (10 $recurrences$) are demonstrated in Fig.~\ref{fig:supfig_scale} (b-e).
We can see that although the overall differences given by different $q$ values are not significant, there are still noteworthy distinctions that confirm the scale-dependent nature of predictability.


\clearpage

\section{Information theory-based extension of $\alpha_\eta$}

\begin{figure}[h!]
\centering
\includegraphics[width=0.9\columnwidth]{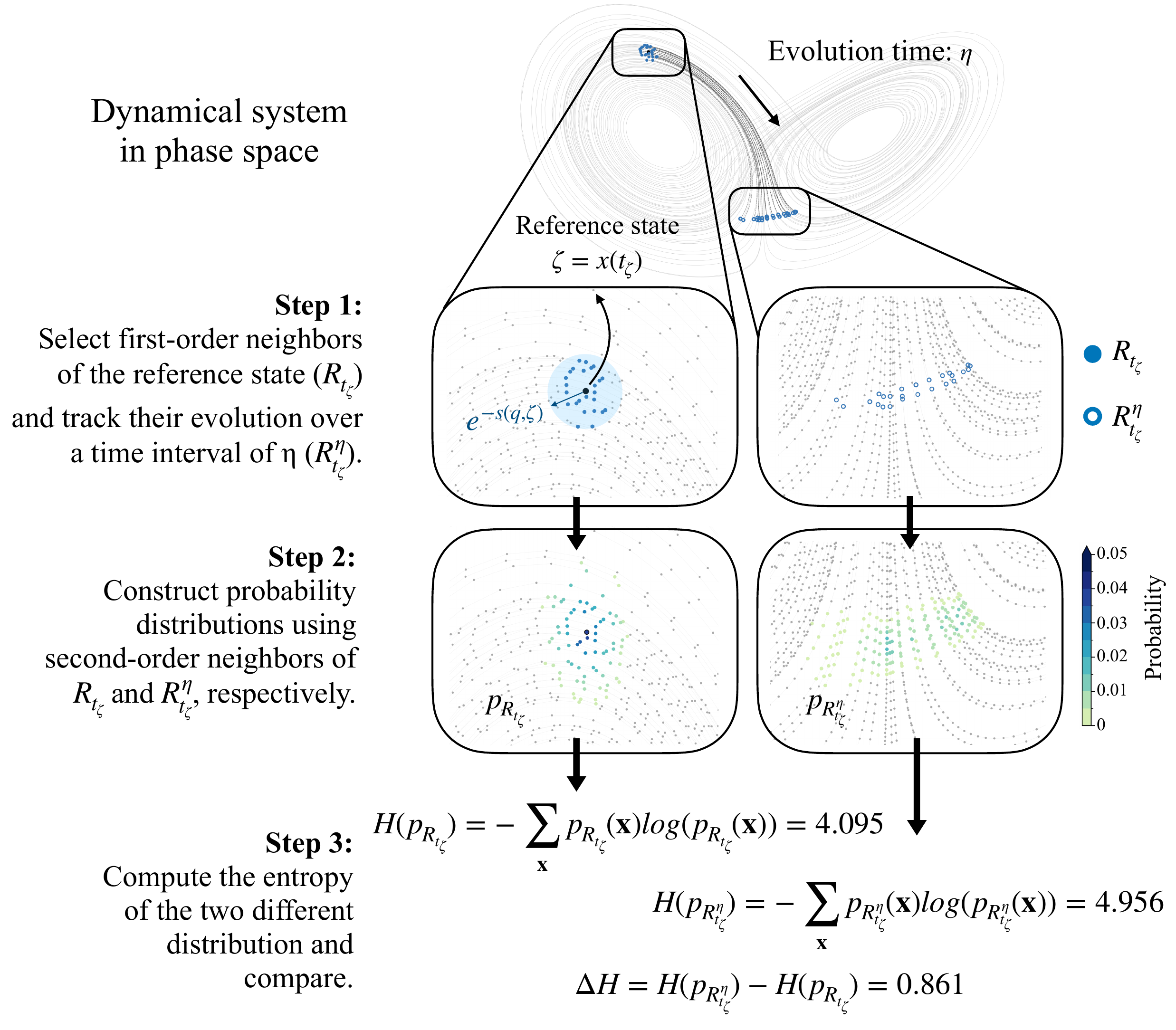}
\caption{Schematic illustration of the computation of entropy change $\Delta H_{\eta}(\zeta)$, demonstrated in the phase space of the Lorenz-63 system. The panels in the second row (step 1) provide a zoomed-in view of the phase space region where $\alpha_{\eta}(\zeta)$ is defined. \textit{Recurrences} ($R_{t_{\zeta}}$) are represented by solid blue dots, while \textit{forward recurrences} ($R_{t_{\zeta}}^{\eta}$) are depicted by empty blue dots, as in Fig. 1. The blue circle with radius $e^{-s(q, \zeta)}$ indicates the hypersphere used to define the neighborhood of the reference state. In step 2, second-order neighbors of each state within $R_{t_{\zeta}}$ and $R_{t_{\zeta}}^{\eta}$ are taken to construct probability distributions $p_{R_{t_{\zeta}}}$ and $p_{R_{t_{\zeta}}^{\eta}}$, respectively. The panels in the third row are colored according to these distributions. In step 3, the entropy of these two distributions is computed and compared.}
\label{fig:information} 
\end{figure}

In this section, we propose a link between the Shannon entropy, and our predictability index $\alpha_{\eta}$. 
Local predictability of a dynamical system can be interpreted as the divergence rate of neighboring trajectories. 
As shown in step 1 of Fig.~\ref{fig:information}, we can adopt the same definitions of $R_{t_{\zeta}}$ and $R_{t_{\zeta}}^{\eta}$ as in $\alpha_\eta$, and use second-order neighbors to characterize the divergence of states within $R_{t_{\zeta}}^{\eta}$ (step 2, in Fig.~\ref{fig:information}). 
Here, second-order neighbors are defined as the recurrences of all states within $R_{t_{\zeta}}$ and $R_{t_{\zeta}}^{\eta}$.

Let us set $R_{t_{\zeta}}$, and recall that it contains $N_{n}$ elements (states). 
Now, let us assume that for each of these elements, $N_{m}$ second-order neighbors are selected, where $N_{m}$ is a predefined parameter (set to $N_{n}$ in this section), resulting in a total of $N_{n} \times N_{m}$ potentially repeated states.
These states can then be viewed as a probability distribution over all the states on the attractor by counting their occurrences and normalizing by $N_{n} \times N_{m}$, resulting in $p_{R_{t_{\zeta}}}$, as shown in Fig.~\ref{fig:information}. 
If the states in $R_{t_{\zeta}}$ are closely located in the phase space, the distribution will be more concentrated; conversely, if they are more spread, the distribution will be less concentrated.
The same steps and analysis also apply to $R_{t_{\zeta}}^{\eta}$.
If we now take the Shannon entropy perspective to characterize the predictability of a state ${\zeta}$, we can compute the rate of change of the entropy over a time interval $\eta$: $\Delta H_{\eta}(\zeta) = H(p_{R_{t_{\zeta}}}) - H(p_{R_{t_{\zeta}}^\eta})$ (step 3, in Fig.~\ref{fig:information}). This quantity measures the expected amount of information created. The more information is created, the more unexpected the configuration, and thus the less predictable the original state. We note that, in the illustrative example presented in Fig.~\ref{fig:information}, $\Delta H_{\eta}(\zeta) > 0$, suggesting that $R_{t_{\zeta}}^{\eta}$ is more spread in the phase space than $R_{t_{\zeta}}$.

We consider this an effective link between $\alpha_\eta$ and information theory, noting that it would require further in-depth analysis, which lies beyond the scope of this paper.
Yet, we believe that the proposed methodology may pave the way for new applications of information theory in high-dimensional dynamical systems.

\clearpage
\section{Comparison with Nonlinear Local Lyapunov Exponent}

\begin{figure}[h!]
\centering
\includegraphics[width=0.8\columnwidth]{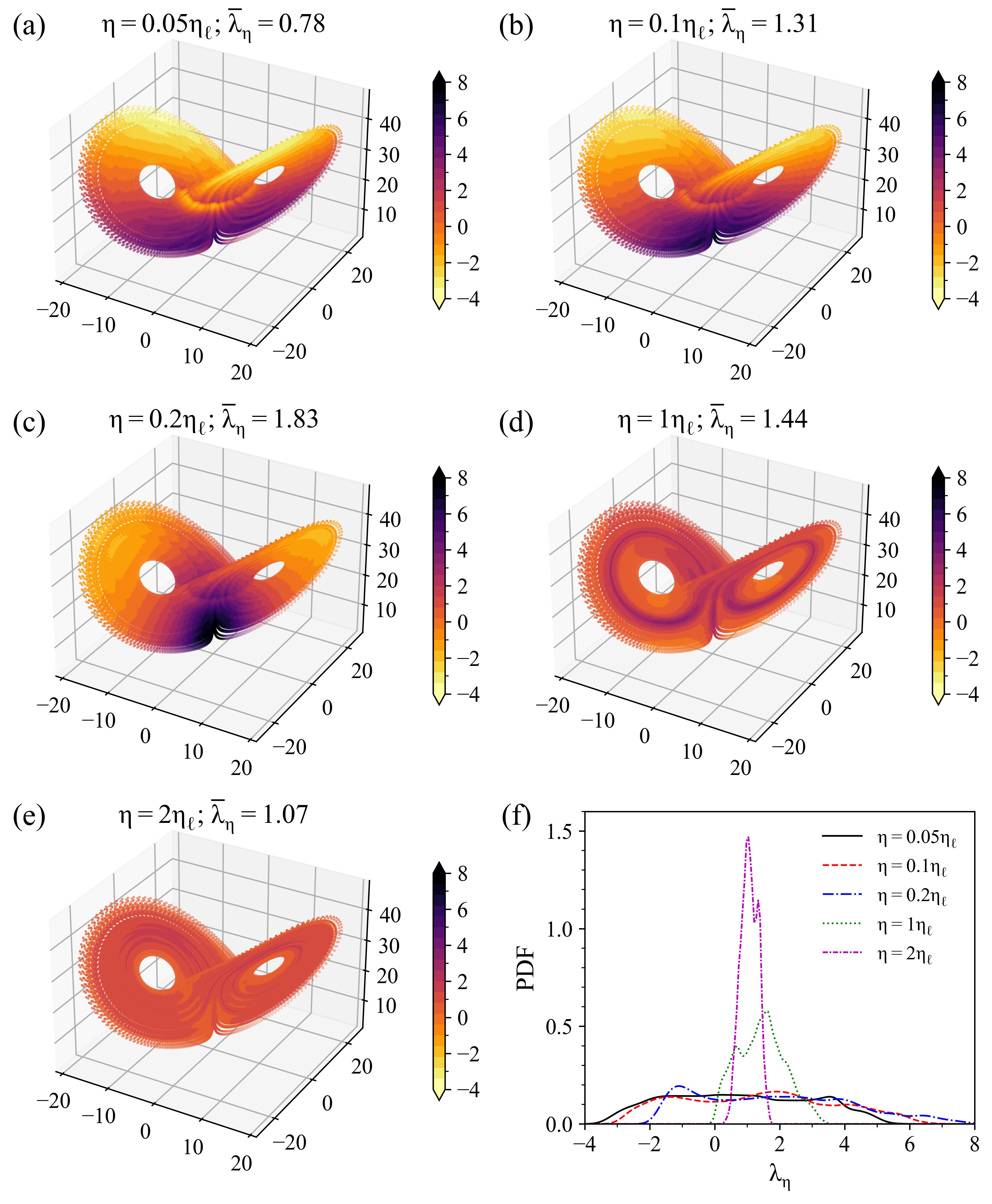}
\caption{Distribution of NLLE at different prediction horizons for the stochastic Lorenz-63 system. Same as Figure 2, but for the Nonlinear Local Lyapunov Exponent.}
\label{fig:supfig_NLLE} 
\end{figure}

The Nonlinear Local Lyapunov Exponent (NLLE) was first proposed by Ding and Li in 2007~\cite{ding2007nonlinear}. Unlike previous approaches that rely on linearized dynamics, NLLE uses the full dynamics to characterize the error growth rate without linearizing the governing equations. When the governing equations of the dynamical system is not explicitly known, they provided a method for estimating NLLE from observational data based on analogous, as shown in~\cite{li2011temporal}. This index is sometimes also called the 'Local Growth Rate', with an open-source implementation provided in~\cite{datseris2022nonlinear}. In this section, we computed NLLE using the same Lorenz-63 dataset as used in the main analysis. Figure~\ref{fig:supfig_NLLE} demonstrates the distribution of NLLE ($\lambda_\eta$) for different $\eta$.

\clearpage
\section{Extension of the Predictability Index}\label{sec:extension}

In this paper, we introduce a new method that characterizes the local predictability of a dynamical system by applying the concept that measuring predictability can be viewed as assessing the divergence rate of nearby trajectories. This is implemented by counting the number of trajectories that evolve closely with the reference trajectory, normalized to a bounded value using the size of the neighborhood. Despite this approach demonstrating potential in analyzing local predictability, it discards information about trajectories that have left the neighborhood, whose dynamics also reflect predictability to some extent. To this end, we proposed an extension of this predictability index by weighting all the nearby trajectories by their distances to the reference trajectory. 

Recall that when deriving $\alpha_\eta$ for a given reference state $\zeta$ and forecasting horizon $\eta$, we defined three groups of states: \textit{recurrences} ($R_{t_{\zeta}}$), \textit{forward recurrences} ($R_{t_{\zeta}}^{\eta}$) and \textit{forward-reference-state recurrences} ($R_{t_{\zeta}+\eta}$). Based on these three groups of states, we define the value of the original version $\alpha_\eta$ as follows:
\begin{equation}
\alpha_{\eta}(\zeta) = \frac{|R_{t_{\zeta}}^\eta \cap R_{{t_{\zeta}+\eta}}|}{|R_{t_{\zeta}}|}
\end{equation}
To extend this index, we first track the distances of all states belonging to \textit{forward recurrences} ($R_{t_{\zeta}}^{\eta}$) to the reference trajectory as $D$:
\begin{equation}
D = \left\{\operatorname{dist}\left({x}(t), {x}({t_{\zeta}+\eta})\right) \;\;\; \forall x(t) \in R_{t_{\zeta}}^{\eta} \right\}.
\end{equation}
where the dist function can be any distance metric, but we only consider L2 norm in this study. 

Next, we also record the distances of states that belong to the intersection of \textit{forward recurrences} ($R_{t_{\zeta}}^{\eta}$) and \textit{forward-reference-state recurrences} ($R_{t_{\zeta}+\eta}$) to the reference trajectory. We denote these distances as $D^\prime$:
\begin{equation}
D^\prime = \left\{\operatorname{dist}\left({x}(t), {x}({t_{\zeta}+\eta})\right) \;\;\; \forall x(t) \in R_{t_{\zeta}}^\eta \cap R_{{t_{\zeta}+\eta}} \right\}.
\end{equation}

Based on these distances, we further define the extension of the index ${\alpha_\eta^\ast}(\zeta)$ as:
\begin{equation}
{\alpha_\eta^\ast}(\zeta) = \frac{\sum_{d^\prime \in D^\prime} {\phi}(d^\prime)}{\sum_{d \in D} {\phi}(d)}
\end{equation}
where $\phi$ can be a custom positive function used to make the extension focus on different aspects, as further explained later. Since $D^\prime$ is a subset of $D$, the values of ${\alpha_\eta^\ast}$, like $\alpha_\eta$, are also defined to lie within the range of 0 to 1. Specifically, we note that the original $\alpha_\eta$ can be viewed as a special case of the extended index ${\alpha_\eta^\ast}$ when $\phi(x) = x^{0}$, meaning that all trajectories are not weighted by their distances to the reference trajectory. 

\begin{figure}[h!]
\centering
\includegraphics[width=0.8\columnwidth]{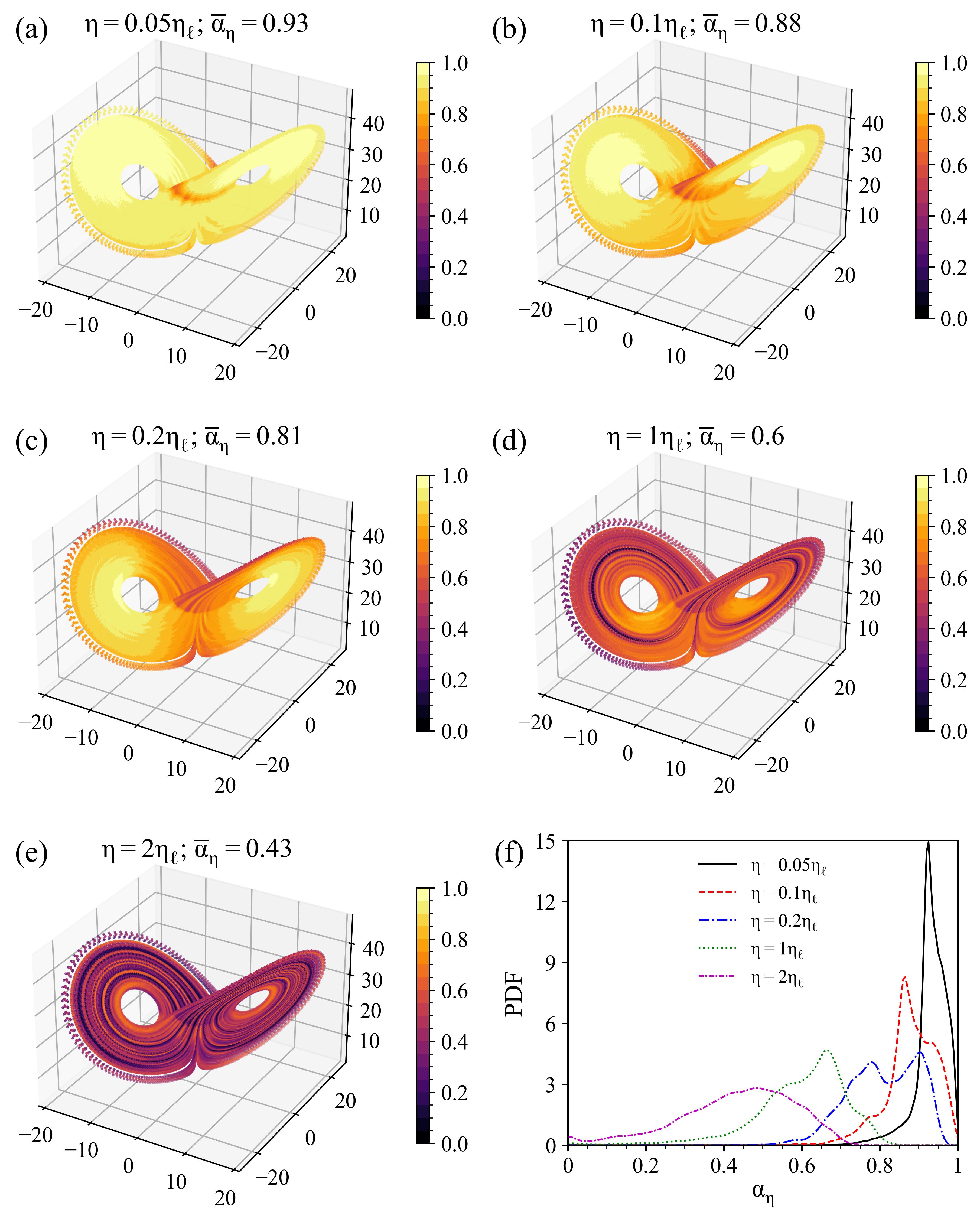}
\caption{Distribution of ${\alpha_\eta^\ast}$ at different prediction horizons for the Lorenz-63 system, with $\phi(x) = x^{-1}$. Same as Figure 2, but for the extension of the predictability index.}
\label{fig:supfig_l_-1} 
\end{figure}

\begin{figure}[h!]
\centering
\includegraphics[width=0.8\columnwidth]{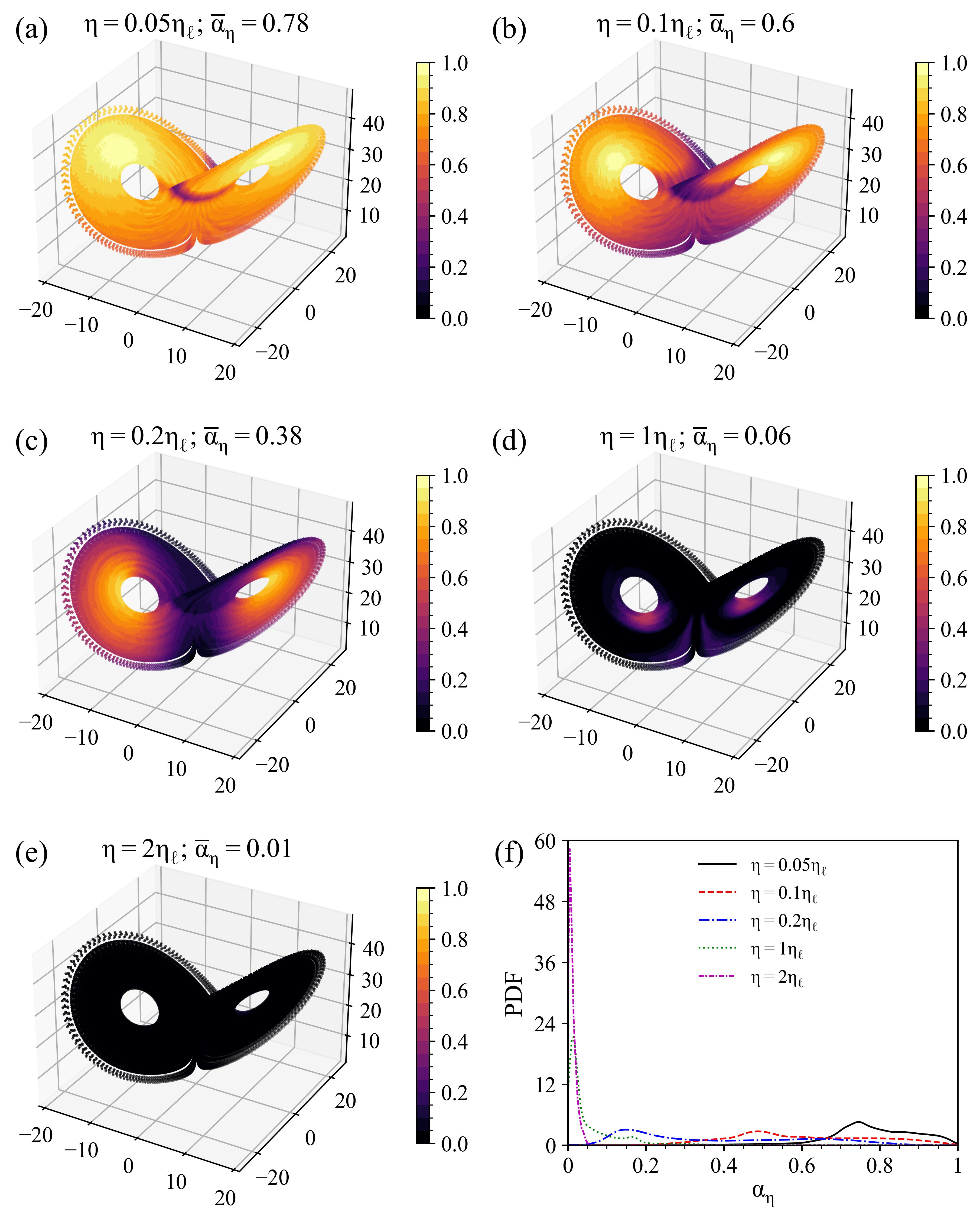}
\caption{Distribution of ${\alpha_\eta^\ast}$ at different prediction horizons for the Lorenz-63 system, with $\phi(x) = x$. Same as Figure 2, but for the extension of the predictability index.}
\label{fig:supfig_l_1} 
\end{figure}

In Figure~\ref{fig:supfig_l_-1} and~\ref{fig:supfig_l_1}, we show the results of ${\alpha_\eta^\ast}$ using $\phi(x) = x^{-1}$ and $\phi(x) = x$, respectively. For these two different $\phi (x)$, the results are generally similar to those obtained when $\phi(x)=1$, although they show noticeable differences in the details. When $\phi(x) = x^{-1}$, ${\alpha_\eta^\ast}$ decreases slowly as the forecasting horizon increases, whereas the opposite holds for $\phi(x) = x^{1}$. This can be explained as $\phi(x) = x^{-1}$ assigns higher weights to the close neighbors, while $\phi(x) = x^{1}$ assigns higher weights to the states that are farther away. This means we can selectively define $\phi$ based on the needs of different tasks. For instance, in some cases, we may be particularly interested in fast diverging trajectories, making $\phi(x) = x^{1}$ a more suitable choice. 

To sum up, this extension uses distances as weights, incorporating more information compared to the original $\alpha_\eta$ which relies on binary values. It also estimates local predictability, and with the extensible nature provided by the weighting function $\phi$, it offers additional information for analysis.


\clearpage
\section{Definition of North Atlantic Wintertime Weather Regimes}

\begin{figure}[h!]
\centering
\includegraphics[width=0.8\columnwidth]{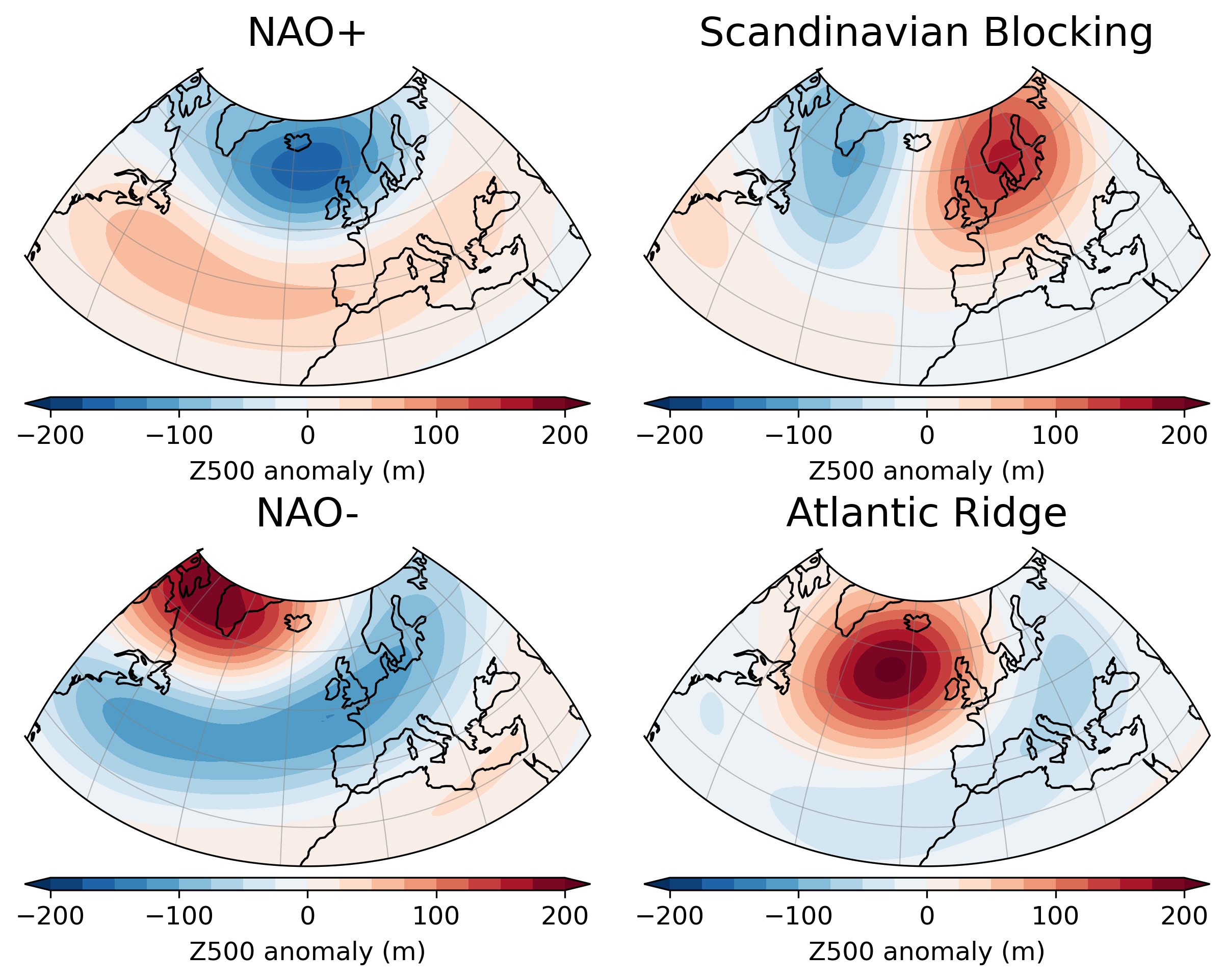}
\caption{Composite Z500 anomaly maps for the four weather regimes in the North Atlantic-Europe region.}
\label{fig:supfig_WR_anomaly}
\end{figure}

The definition of weather regimes is adapted from~\cite{cassou2008intraseasonal}, focusing on the wintertime (DJFM) 500 hPa (Z500) in the Euro-Atlantic sector (80°W to 50°E, 22.5°N to 70°N). The weighted Z500 daily anomaly data is decomposed using EOF analysis, retaining the 14 leading modes (accounting for approximately 90\% of the variance) to speed up the computation. The climatology is computed as the calendar day mean field of Z500 from 1979 to 2022, based on a centered 15-day window, while the weight is applied to the data as a cosine function of the latitude. The k-means clustering algorithm is then applied to the principal components, resulting in four weather regimes, whose centroids are demonstrated in Fig.~\ref{fig:supfig_WR_anomaly}. Next, we compute the normalized projection of each daily Z500 anomaly field onto the cluster centroid, which has been termed the Weather Regimes Index (IWR) in previous studies~\cite{michel2011link}. Here, we aim to analyze the predictability of established weather regimes; therefore, we require that the Weather Regimes Index (IWR) remains above 1.0 for at least 5 days to classify a state as belonging to a weather regime. States that are not allocated to any weather regime are classified as "no regime".

\clearpage
\section{Datesets for other examples}

\subsection{Rössler system}

The Rössler system is a set of three nonlinear ordinary differential equations originally introduced in 1976~\cite{rossler1976equation}. Its governing equations can be written as:

\begin{eqnarray}
\frac{dx}{dt} &=& -y-z, \\
\frac{dy}{dt} &=& x + ay, \\
\frac{dz}{dt} &=& b + z(x-c).
\end{eqnarray}  

In this study, we adopt the classical parameters of the Rössler system that exhibit deterministic chaos, where $(a,b,c)=(0.2,0.2,5.7)$. Observational data are obtained through numerical simulation with a temporal resolution of $dt = 0.005$ over $10^5$ time steps. 

\subsection{Spring-slider system for laboratory earthquakes}

In~\cite{gualandi2023deterministic}, one spring-slider model is developed to simulate the dynamics of slow earthquakes. This model is described by the following set of ordinary differential equations (ODEs):

\begin{eqnarray}
\dot{x} &=& \frac{e^x\left[\left(\beta_1-1\right) x(1+\lambda u)+y-u\right]+\kappa\left(\frac{v_0}{v_*}-e^x\right)-\dot{u} \frac{1+\lambda y}{1+\lambda u}}{1+\lambda u+v e^x} \\
\dot{y} &=& \kappa\left(\frac{v_0}{v_*}-e^x\right)-v e^x \dot{x} \\
\dot{z} &=& -\rho e^x\left(\beta_2 x+z\right) \\
\dot{u} &=& -\alpha-\gamma u+\dot{z}.
\end{eqnarray}  
where the state of system can be fully characterized by the state vector $\left[x, y, z, u\right]$. To further simulate the irregularity of labquakes, two stochastic terms modeled by a Wiener process are added to the equations for $y$ and $u$, making the system a set of Stochastic Differential Equations (SDE):

\begin{eqnarray}
\mathrm{d} y &=& \left[\kappa\left(\frac{v_0}{v_*}-e^x\right)-v e^x \dot{x}\right] \mathrm{d} T+\varepsilon_y \mathrm{~d} W_T \\
\mathrm{~d} u &=& [-\alpha-\gamma u+\dot{z}] \mathrm{d} T+\varepsilon_u \mathrm{~d} W_T
\end{eqnarray}

The values of all the parameters are presented in Table~\ref{tab:parameter}, while their physical explanations are available in~\cite{gualandi2023deterministic}. We generated $2*10^5$ time steps with $dt = 0.01$ for the analysis. In Figure 7, the attractor for the spring-slider system is visualized only only using variables x, y, and z, given the limitation to three dimensions.

\begin{table}[h]
    \centering
    \caption{Simulation Parameters}
    \begin{tabular}{ll}
        \toprule
        Parameter & Value \\ 
        \midrule
        $\beta_1$ & $1.2$ \\
        $\beta_2$ & $0.8265$ \\
        $\lambda$ & $0.0156$ \\
        $\kappa$ & $0.2555$ \\
        $v_0$ & $10^{-6}$ \\
        $v_*$ & $10^{-6}$ \\
        $v$ & $0.004$ \\
        $\rho$ & $0.1$ \\
        $\alpha$ & $0.0112$ \\
        $\gamma$ & $0.03$ \\
        $\varepsilon_y$ & $0.0115$ \\
        $\varepsilon_u$ & $0.0111$ \\
        \bottomrule
    \end{tabular}
    \label{tab:parameter}
\end{table}

\subsection{Double Pendulum, EEG and ECG}

The data of these three systems are from~\cite{brunton2017chaos}, with the double pendulum data
generated through numerical simulation, while the other two are obtained from observational biological data. The EEG and ECG data were adapted from the PhysioNet database~\cite{goldberger2000physiobank,laguna1997database,kemp2000analysis}.

For the double pendulum, the simulations were conducted using a variational integrator based on the Euler-Lagrange equations:
\begin{equation}    
\frac{\mathrm{d}}{\mathrm{d}t} \frac{\partial L}{\partial \dot{\mathbf{q}}} - \frac{\partial L}{\partial \mathbf{q}} = 0,
\end{equation}
with the Lagrangian $L = T - V$ is the kinetic (T) minus potential (V) energy. For double pendulum, $\mathbf{q}=\left[\theta_1 \quad \theta_2\right]^T$, and the Lagrangian (L) becomes:
\begin{equation} 
L = \frac{1}{2}(m_1 + m_2)l_1^2\dot{\theta}_1^2 + \frac{1}{2}m_2l_2^2\dot{\theta}_2^2 + m_2l_1l_2\dot{\theta}_1\dot{\theta}_2\cos(\theta_1 - \theta_2) - (m_1 + m_2)l_1g(1 - \cos\theta_1) - m_2l_2g(1 - \cos\theta_2).
\end{equation} 
where $l_1 = l_2 = m_1 = m_2 = 1$, $g = 10$. The action integral was approximated using the trapezoidal rule to compute the equations of motion:

\begin{equation} 
\delta \int_a^b L(\mathbf{q}, \dot{\mathbf{q}}, t) \, \mathrm{d}t = 0.
\end{equation} 

$x(t) = \sin(\theta_1)$ is used as the observational time series for double pendulum.

For all these three examples, a similar approach to~\cite{brunton2017chaos} is applied to the time series data to obtain eigen-time-delay coordinates. Specifically, a time series x(t) is stacked into a Hankel matrix H, which is then decomposed using singular value decomposition (SVD):

\begin{equation}
\mathbf{H}=\left[\begin{array}{cccc}
x\left(t_1\right) & x\left(t_2\right) & \cdots & x\left(t_p\right) \\
x\left(t_2\right) & x\left(t_3\right) & \cdots & x\left(t_{p+1}\right) \\
\vdots & \vdots & \ddots & \vdots \\
x\left(t_q\right) & x\left(t_{q+1}\right) & \cdots & x\left(t_m\right)
\end{array}\right]=\mathbf{U} \Sigma \mathbf{V}^* .
\end{equation}

This yields a $\mathbf{V}^*$ matrix, which represents a hierarchy of eigen time series that reconstruct a delay-embedded attractor. The predictability index $\alpha_\eta$ is computed based on the $\mathbf{V}^*$ matrix and visualized on the attractor constructed from the three leading time series in $\mathbf{V}^*$ (see Fig. 7). A summary of the parameters used for the construction and decomposition of the Hankel matrix is provided in Table~\ref{tab:summary}. We note that the parameters used here differ from those in~\cite{brunton2017chaos} to retain higher energy for predictability analysis.

\begin{table}[h!]
\centering
\caption{Summary of Parameters}
\begin{tabular}{@{}llllllll@{}}
\toprule
\textbf{System}       & \textbf{Type} & \textbf{Measured Variable}  & \textbf{Time Step} $\Delta t$ & \textbf{\# Rows in H} $q$ & \textbf{Rank} $r$ & \textbf{Energy in $r$ Modes (\%)} \\ \midrule
Double Pendulum         & ODE           & $x(t)$                   & 0.001                         & 100                        & 5                   & 99.9                                  \\
ECG               & Data           & Voltage                  & 0.004s                         & 25                        & 5                    & 93.3                           \\
EEG          & Data           & Voltage                   & 0.01s                         & 200                        & 8                    & 61.2                              \\ \bottomrule
\end{tabular}
\label{tab:summary}
\end{table}

\clearpage
\section{Real-time Predictability Analysis}\label{sec:realtime}

In certain scenarios, such as weather forecasting, real-time predictability analysis is important for evaluating and informing the operational forecast system.
However, the computation of $\alpha_\eta$ rely on the future trajectory of the reference state $\zeta^{\prime}$, therefore hindering direct real-time applications. 
As shown in Equation 9 in the main text, we propose a proxy that do not rely on the future information of the reference state. 
The idea of this proxy stems from the fact that the predictability index represents local predictability in the phase space. Therefore, we can use the average predictability of the neighboring states of $\zeta^{\prime}$ to represent its predictability.

\begin{figure}[h!]
\centering
\includegraphics[width=0.66\columnwidth]{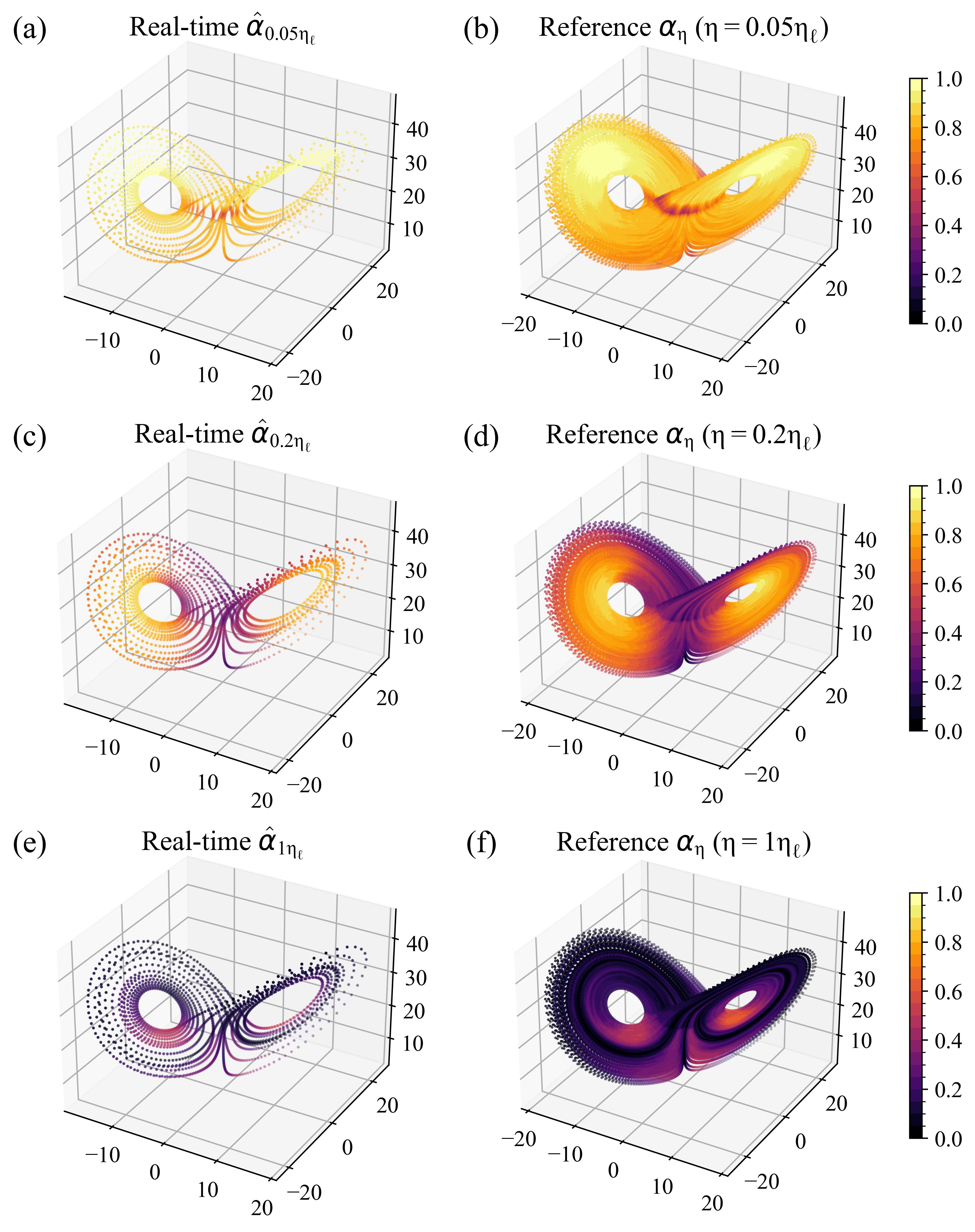}
\caption{Real-time predictability analyses for the Lorenz system. (a,c,e) 2000 unseen Lorenz-63 states colored according to the predictability index proxy $\hat{\alpha_{\eta}}$, presented across three different forecasting horizons: 0.05$\eta_{\ell}$ (a), 0.2$\eta_{\ell}$ (c), and 1$\eta_{\ell}$ (e). Panels (b), (d), and (f) display the reference $\alpha_{\eta}$, as in 2 (a), (c), and (d).}
\label{fig:realtime}  
\end{figure}

An example of a real-time application is presented using the Lorenz-63 system, following the same numerical simulation scheme as other experiments in this study. We assume that we have $10^5$ time steps as historical observations, and their corresponding $\alpha_\eta$ have already been computed, as shown in Fig.~\ref{fig:realtime} (b,d,f). 
Meanwhile, we also assume there are 2000 states not present in the historical observations, which are used to simulate real-time unseen states in an operational forecast system.
Figure~\ref{fig:realtime} (a,c,e) shows the predictability proxy for the unseen states. By comparing the results with the reference $\alpha_\eta$, we note that the proposed proxy can effectively capture the features of predictability, regardless of the region in phase space or the forecasting horizon. 
Specifically, the low predictability near the lobe transition region is in Fig.~\ref{fig:realtime} (a,c), and the high predictability region near the fixed point in Fig.~\ref{fig:realtime} (e) are both well reflected. Although this application is a rather simple case, we anticipate its application in other operational forecast systems.

\clearpage
\end{document}